%% file: ms.tex
\documentclass[conference, 10pt]{IEEEtran}
\input{sec-preamble.tex}

\usepackage[T1]{fontenc}
\newcommand\thefont{\expandafter\string\the\font}
\usepackage{enumitem}
\usepackage{pgf}
\usepackage{svg}
 
\usepackage{pgfplots}
\pgfplotsset{compat=newest}
\usetikzlibrary{plotmarks}
\usetikzlibrary{arrows.meta}
\usepgfplotslibrary{patchplots}
\usepackage{grffile}
\pgfplotsset{plot coordinates/math parser=false}

\begin{document}
\input{sec-title.tex}
\input{sec-abstract.tex}
\input{sec-keywords.tex}

\glsresetall

\input{sec-introduction.tex}
\input{sec-system-model.tex}
\input{sec-contribution.tex}
\input{sec-numerical-results.tex}
\input{sec-conclusion.tex}
\input{sec-acknowledgement.tex}
\input{sec-bibliography.tex}

\end{document}

%% file: sec-preamble.tex
\usepackage{booktabs}
\usepackage{array}

\usepackage{cite}
\ifCLASSINFOpdf
  \usepackage[pdftex]{graphicx}
  \usepackage{epstopdf} % remove this
\else
\fi
\usepackage{amsmath}
\usepackage{algorithm}
\usepackage{algorithmic}

\usepackage[caption=false,font=footnotesize]{subfig}
\usepackage{url}
\usepackage{amsthm}

\newtheoremstyle{mydefinition}% name of the style to be used
{}% measure of space to leave above the theorem. E.g.: 3pt
{}% measure of space to leave below the theorem. E.g.: 3pt
{}% name of font to use in the body of the theorem
{0pt}% measure of space to indent
{\bfseries}% name of head font
{.}% punctuation between head and body
{ }% space after theorem head; " " = normal interword space
{\thmname{#1}\thmnumber{ #2}: \thmnote{#3}}

\theoremstyle{mydefinition}

\newtheoremstyle{myremark}% name of the style to be used
{}% measure of space to leave above the theorem. E.g.: 3pt
{}% measure of space to leave below the theorem. E.g.: 3pt
{}% name of font to use in the body of the theorem
{0pt}% measure of space to indent
{\bfseries}% name of head font
{.}% punctuation between head and body
{ }% space after theorem head; " " = normal interword space
{\thmname{#1}\thmnumber{ #2}: \thmnote{#3}}
\theoremstyle{myremark}

\newtheoremstyle{remarkshort}% name of the style to be used
{}% measure of space to leave above the theorem. E.g.: 3pt
{}% measure of space to leave below the theorem. E.g.: 3pt
{}% name of font to use in the body of the theorem
{0pt}% measure of space to indent
{\bfseries}% name of head font
{.}% punctuation between head and body
{ }% space after theorem head; " " = normal interword space
{\thmname{#1}\thmnumber{ #2}}

\theoremstyle{remarkshort}

\theoremstyle{remarkshort}

\input{math-ipr.tex}
\input{glossary.tex}

\newcommand{\tabref}[1]{Table~\ref{#1}}
\newcommand{\figref}[1]{\figurename~\ref{#1}}

\usepackage{mathtools}
\usepackage{comment}

%% file: math-ipr.tex
\usepackage{amssymb,amsmath}
\usepackage{xspace}
\usepackage{bm}
\usepackage{bbm}
\usepackage{physics}

\newcommand{\los}{\mathsf{LOS}}
\newcommand{\nlos}{\mathsf{NLOS}}

\newcommand{\set}[1]{\ensuremath{\mathcal{#1}}\xspace} % caligraphic set notation
\newcommand{\mat}[1]{\ensuremath{\mathbf{#1}}\xspace} % matrices
\newcommand{\parens}[1]{{\left(#1\right)}\xspace}
\newcommand{\brackets}[1]{{\left[#1\right]}\xspace}
\renewcommand{\braces}[1]{{\left\{#1\right\}}\xspace}

\renewcommand{\real}{{\mathbb{R}}\xspace}
\renewcommand{\imaginary}{{\mathbb{I}}\xspace}
\newcommand{\complex}{{\mathbb{C}}\xspace}

\renewcommand{\j}{{\mathrm{j}}}

\newcommand{\setcomplex}{\ensuremath{\complex}}

\newcommand{\setvector}[2]{\ensuremath{#1^{#2 \times 1}}\xspace}

\newcommand{\setvectorcomplex}[1]{\setvector{\setcomplex}{#1}}

\newcommand{\setmatrix}[3]{\ensuremath{#1^{#2 \times #3}}\xspace}

\newcommand{\setmatrixcomplex}[2]{\setmatrix{\setcomplex}{#1}{#2}}

\newcommand{\trans}{^{\mathsf{T}}}
\newcommand{\ctrans}{^{{*}}\xspace}
\newcommand{\logtwo}[1]{{\mathrm{log}_{2}\parens{#1}}}

\newcommand{\diag}[1]{{\mathrm{diag}\parens{#1}}}

\renewcommand{\norm}[1]{\left\Vert{#1}\right\Vert}
\newcommand{\pnorm}[2]{\norm{#2}_{#1}}

\newcommand{\normtwo}[1]{\pnorm{2}{#1}}

\newcommand{\distcgauss}[2]{{\mathcal{N}_{\complex}\parens{#1,#2}}\xspace} % complex Gaussian 
\DeclareMathOperator*{\expect}{\mathbb{E}\xspace}

\newcommand{\subjectto}{\ensuremath{\mathrm{s.t.~}}\xspace}
\newcommand{\opt}{\ensuremath{^{\star}}\xspace}

\newcommand{\pnoise}{\sigma^2}

\newcommand{\snr}{\ensuremath{\mathsf{SNR}}\xspace}

\newcommand{\sinr}{\ensuremath{\mathsf{SINR}}\xspace}

\newcommand{\inr}{\ensuremath{\mathsf{INR}}\xspace}

\newcommand{\Nt}{{N_\mathrm{t}}\xspace} % number of antennas at Tx
\newcommand{\Nr}{{N_\mathrm{r}}\xspace} % number of antennas at Rx

\newcommand{\labeldl}{\mathsf{DL}}
\newcommand{\labelul}{\mathsf{UL}}

\newcommand{\ul}{\labelul}
\newcommand{\dl}{\labeldl}

\def\vf{{\mat{f}}}

\def\vh{{\mat{h}}}

\def\vn{{\mat{n}}}

\def\vw{{\mat{w}}}

\def\vy{{\mat{y}}}
\def\vz{{\mat{z}}}

\def\mF{{\mat{F}}}

\def\mH{{\mat{H}}}
\def\mI{{\mat{I}}}

\def\mN{{\mat{N}}}

\def\mW{{\mat{W}}}

\def\sD{{\set{D}}}

%% file: glossary.tex
\usepackage{xspace}
\usepackage[acronym,nogroupskip,nonumberlist,nopostdot]{glossaries}
\usepackage{xcolor}

\newacronym{snr}{SNR}{signal-to-noise ratio}
\newacronym{sinr}{SINR}{signal-to-interference-plus-noise ratio}
\newacronym{inr}{INR}{interference-to-noise ratio}
\newacronym{sir}{SIR}{signal-to-interference ratio}
\newacronym{sqr}{SQR}{signal-to-quantization-noise ratio}
\newacronym{sqnr}{SQNR}{signal-to-quantization-plus-noise ratio}
\newacronym{ian}{IAN}{interference as noise}
\newacronym{ber}{BER}{bit error rate}
\newacronym{pn}{PN}{pseudorandom noise}
\newacronym{bfsk}{BFSK}{binary frequency shift keying}
\newacronym{fh}{FH}{frequency-hopped}
\newacronym{fh-bfsk}{FH-BFSK}{frequency-hopped binary frequency shift keying}
\newacronym{crc}{CRC}{cyclic redundancy check}
\newacronym{isi}{ISI}{intersymbol interference}
\newacronym{dsss}{DSSS}{direct-sequence spread spectrum}
\newacronym{ofdm}{OFDM}{orthogonal frequency-division multiplexing}
\newacronym{ofdma}{OFDMA}{orthogonal frequency-division multiple access}
\newacronym{sdr}{SDR}{software-defined radio}
\newacronym{tx}{TX}{transmitter}
\newacronym{rx}{RX}{receiver}
\newacronym{fdd}{FDD}{frequency-division duplexing}
\newacronym{tdd}{TDD}{time-division duplexing}
\newacronym{fdma}{FDMA}{frequency-division multiple access}
\newacronym{tdma}{TDMA}{time-division multiple access}
\newacronym{sdma}{SDMA}{space-division multiple access}
\newacronym[plural=MPCs]{mpc}{MPC}{multipath component}
\newacronym{mui}{MUI}{multi-user interference}
\newacronym{lsb}{LSB}{least significant bit}
\newacronym{jcas}{JCAS}{joint communication and sensing}
\newacronym{qam}{QAM}{quadrature amplitude modulation}
\newacronym{mqam}{MQAM}{M-ary quadrature amplitude modulation}
\newacronym{mrt}{MRT}{maximum ratio transmission}
\newacronym{mrc}{MRC}{maximum ratio combining}

\newacronym{ls}{LS}{least-squares}
\newacronym{lms}{LMS}{least mean squares}
\newacronym{rls}{RLS}{recursive least-squares}
\newacronym{rzf}{RZF}{regularized zero-forcing}
\newacronym{mmse}{MMSE}{minimum mean square error}
\newacronym{lmmse}{LMMSE}{linear minimum mean square error}
\newacronym{mse}{MSE}{mean square error}
\newacronym{fft}{FFT}{fast Fourier transform}
\newacronym{dft}{DFT}{discrete Fourier transform}
\newacronym{dtft}{DTFT}{discrete-time Fourier transform}
\newacronym{ctft}{CTFT}{continuous-time Fourier transform}
\newacronym{ml}{ML}{machine learning}
\newacronym{dlr}{DL}{deep learning}
\newacronym{nn}{NN}{neural network}
\newacronym[plural=RNNs]{rnn}{RNN}{recurrent neural network}
\newacronym[plural=ADCs]{adc}{ADC}{analog-to-digital converter}
\newacronym[plural=DACs]{dac}{DAC}{digital-to-analog converter}
\newacronym[plural=FPGAs]{fpga}{FPGA}{field-programmable gate array}
\newacronym{evm}{EVM}{error vector magnitude}
\newacronym{enob}{ENOB}{effective number of bits}
\newacronym{zf}{ZF}{zero-forcing}
\newacronym{rv}{r.v.}{random variable}
\newacronym{omp}{OMP}{orthogonal matching pursuit}
\newacronym{svd}{SVD}{singular value decomposition}

\newacronym{sdp}{SDP}{semidefinite programming}
\newacronym{psd}{PSD}{positive semidefinite}
\newacronym{nsd}{NSD}{negative semidefinite}

\newacronym{ks}{K-S}{Kolmogorov-Smirnov}
\newacronym{mad}{MAD}{median absolute deviation around the median}
\newacronym{awgn}{AWGN}{additive white Gaussian noise}

\newacronym{agc}{AGC}{automatic gain control}
\newacronym{rf}{RF}{radio frequency}
\newacronym{if}{IF}{intermediate frequency}
\newacronym{los}{LOS}{line-of-sight}
\newacronym{nlos}{NLOS}{non-line-of-sight}
\newacronym{ple}{PLE}{path loss exponent}
\newacronym[plural=dB,firstplural=decibels (dB)]{db}{dB}{decibel}
\newacronym[plural=dBm,firstplural=decibel milliwatts (dBm)]{dbm}{dBm}{decibel milliwatts}
\newacronym{pa}{PA}{power amplifier}
\newacronym{lna}{LNA}{low noise amplifier}
\newacronym{vga}{VGA}{variable gain amplifier}
\newacronym{cw}{CW}{continuous wave}
\newacronym{papr}{PAPR}{peak-to-average power ratio}
\newacronym{usrp}{USRP}{Universal Software Radio Peripheral}
\newacronym{irr}{IRR}{image rejection ratio}
\newacronym{lo}{LO}{local oscillator}
\newacronym{vm}{VM}{vector modulator}
\newacronym{mmwave}{mmWave}{millimeter-wave}
\newacronym{eirp}{EIRP}{effective isotropic radiated power}
\newacronym{rsrp}{RSRP}{reference signal received power}

\newacronym{csma}{CSMA}{carrier-sense multiple access}
\newacronym{csmaca}{CSMA/CA}{carrier-sense multiple access with collision avoidance}
\newacronym{csmacd}{CSMA/CD}{carrier-sense multiple access with collision detection}
\newacronym{mac}{MAC}{medium access control}
\newacronym{phy}{PHY}{physical layer}
\newacronym{4g}{4G}{fourth generation}
\newacronym{lte}{LTE}{Long-Term Evolution}
\newacronym{4glte}{4G LTE}{\gls{4g} \gls{lte}}
\newacronym{5g}{5G}{fifth generation}
\newacronym{nr}{NR}{New Radio}
\newacronym{5gnr}{5G NR}{5G New Radio}
\newacronym{ieee}{IEEE}{Institute of Electrical and Electronics Engineers}
\newacronym{wifi}{Wi-Fi}{IEEE 802.11}
\newacronym{lan}{LAN}{local area network}
\newacronym{wlan}{WLAN}{wireless local area network}
\newacronym[plural=BSs]{bs}{BS}{base station}
\newacronym[plural=SBSs]{sbs}{SBS}{small-cell base station}
\newacronym[plural=FD-SBSs]{fdsbs}{FD-SBS}{\gls{fd}-enabled \gls{sbs}}
\newacronym[plural=MBSs]{mbs}{MBS}{macrocell base station}
\newacronym[plural=UEs]{ue}{UE}{user equipment}
\newacronym{ul}{UL}{uplink}
\newacronym{dl}{DL}{downlink}
\newacronym{qos}{QoS}{Quality of Service}
\newacronym{fcc}{FCC}{Federal Communications Commission}
\newacronym{iab}{IAB}{integrated access and backhaul}
\newacronym{fab}{FAB}{fixed access and backhaul}
\newacronym{hetnet}{HetNet}{heterogeneous network}

\newacronym{siso}{SISO}{single-input single-output}
\newacronym{mimo}{MIMO}{multiple-input multiple-output}
\newacronym{sumimo}{SU-MIMO}{single-user \gls{mimo}}
\newacronym{mumimo}{MU-MIMO}{multi-user \gls{mimo}}
\newacronym{bf}{BF}{beamforming}
\newacronym{ca}{CA}{constant amplitude}
\newacronym{ula}{ULA}{uniform linear array}
\newacronym{upa}{UPA}{uniform planar array}
\newacronym[\glslongpluralkey={angles of arrival}]{aoa}{AoA}{angle of arrival}
\newacronym[\glslongpluralkey={angles of departure}]{aod}{AoD}{angle of departure}
\newacronym{dof}{DoF}{degrees of freedom}
\newacronym{csi}{CSI}{channel state information}
\newacronym{csit}{CSIT}{\gls{csi} at the transmitter}
\newacronym{csir}{CSIR}{\gls{csi} at the receiver}
\newacronym{cs}{CS}{compressed sensing}

\newacronym{fd}{full-duplex}{in-band full-duplex}
\newacronym{hd}{HD}{half-duplex}
\newacronym{si}{SI}{self-interference}
\newacronym{sic}{SIC}{self-interference cancellation}
\newacronym{soi}{SoI}{signal of interest}
\newacronym{asic}{A-SIC}{analog \acrlong{sic}}
\newacronym{dsic}{D-SIC}{digital \gls{sic}}
\newacronym{star}{STAR}{simultaneous transmit and receive}
\newacronym{warp}{WARP}{Wireless Open-Access Research Platform}
\newacronym{bfc}{BFC}{beamforming cancellation}
\newacronym{ipi}{IPI}{inter-panel-interference}
\newacronym{ipic}{IPIC}{inter-panel-interference cancellation}
\newacronym{sse}{SSE}{sum spectral efficiency}

\newacronym{lstm}{LSTM}{long short-term memory}

\newacronym{qcqp}{QCQP}{quadratically-constrained quadratic programming}
\newacronym{pdf}{PDF}{probability density function}
\newacronym{cdf}{CDF}{cumulative distribution function}
\newacronym{iid}{i.i.d.}{independently and identically distributed}

\newacronym{elf}{ELF}{extremely low frequency}
\newacronym{slf}{SLF}{super low frequency}
\newacronym{ulf}{ULF}{ultra low frequency}
\newacronym{vlf}{VLF}{very low frequency}
\newacronym{lf}{LF}{low frequency}
\newacronym{mf}{MF}{medium frequency}
\newacronym{hf}{HF}{high frequency}
\newacronym{vhf}{VHF}{very high frequency}
\newacronym{uhf}{UHF}{ultra high frequency}
\newacronym{shf}{SHF}{super high frequency}
\newacronym{ehf}{EHF}{extremely high frequency}
\newacronym{thf}{THF}{tremendously high frequency}

\newacronym{wncg}{WNCG}{Wireless Networking and Communications Group}
\newacronym{linc}{LINC}{Laboratory of Informatics, Networks, and Communications}
\newacronym{ut}{UT Austin}{The University of Texas at Austin}
\newacronym{uiuc}{UIUC}{University of Illinois at Urbana-Champaign}
\newacronym{usc}{USC}{University of Southern California}
\newacronym{mit}{MIT}{Massachusetts Institute of Technology}
\newacronym{berkeley}{UC Berkeley}{University of California, Berkeley}
\newacronym{osu}{OSU}{Ohio State University}

\newcommand{\mmwave}{\gls{mmwave}\xspace}
\newcommand{\MIMO}{\gls{mimo}\xspace}

\newcommand{\rf}{\gls{rf}\xspace}
\newcommand{\RX}{\acrlong{rx}\xspace}
\newcommand{\TX}{\acrlong{tx}\xspace}

\newcommand{\fd}{\acrshort{fd}\xspace}
\newcommand{\hd}{\acrlong{hd}\xspace}
\newcommand{\DL}{\acrlong{dl}\xspace}
\newcommand{\UL}{\acrlong{ul}\xspace}
\newcommand{\SI}{\acrlong{si}\xspace}

\newcommand{\BS}{\acrlong{bs}\xspace}
\newcommand{\BSs}{\acrlong{bs}s\xspace}
\newcommand{\SNR}{\acrshort{snr}\xspace}
\newcommand{\INR}{\gls{inr}\xspace}
\newcommand{\SINR}{\gls{sinr}\xspace}

\newcommand{\gpsnr}{\glspl{snr}\xspace}

\newcommand{\NN}{\acrlong{nn}\xspace}
\newcommand{\LOS}{\gls{los}\xspace}
\newcommand{\NLOS}{\gls{nlos}\xspace}

\newcommand{\sse}{\acrlong{sse}\xspace}

%% file: sec-title.tex
\title{Site-Specific Beam Learning for Full-Duplex Massive MIMO Wireless Systems}

\author{
\IEEEauthorblockN{Samuel Li and Ian P.~Roberts}%
\IEEEauthorblockA{Wireless Lab, Department of Electrical and Computer Engineering\\University of California, Los Angeles (UCLA), Los Angeles, CA USA}%
\IEEEauthorblockA{Email: \{samuel.li, ianroberts\}@ucla.edu}%
}

\maketitle

%% file: sec-abstract.tex
\begin{abstract}
    Existing beamforming-based \fd solutions for multi-antenna wireless systems often rely on explicit estimation of the \SI channel.
    The pilot overhead of such estimation, however, can be prohibitively high in \acrlong{mmwave} and massive MIMO systems, thus limiting the practicality of existing solutions, especially in fast-fading conditions.
    In this work, we present a novel beam learning framework that bypasses explicit \SI channel estimation by designing beam codebooks to efficiently obtain implicit channel knowledge that can then be processed by a deep learning network to synthesize transmit and receive beams for \fd operation.
    Simulation results using ray-tracing illustrate that our proposed technique can allow a \fd base station to craft serving beams that couple low self-interference while delivering high SNR, with 75--97\% fewer measurements than would be required for explicit estimation of the self-interference channel.
\end{abstract}

%% file: sec-keywords.tex

%% file: sec-introduction.tex
\section{Introduction} \label{sec:introduction}

In-band full-duplex operation in future \BSs has the potential to boost spectral efficiency, reduce latency, and enable advanced sensing capabilities, paving the way towards 6G wireless networks \cite{smida_fd_6g_jsac_2023}.
Realizing such full-duplex \BSs requires effective suppression of the \textit{\SI} that arises when attempting to transmit and receive at the same time across the same frequency band,
as the \BS's \UL signal could otherwise be severely degraded by its own \DL transmission and render \fd infeasible.
In traditional sub-6~GHz wireless systems, \SI cancellation has conventionally been performed through a combination of analog and digital cancellation techniques~\cite{sabharwal_-band_2014,kolodziej_techniques_2019,namyoon_learning_sic_2024}.
However, adapting these methods to \mmwave and massive \MIMO \BSs faces significant challenges due to their large antenna arrays, high operating frequencies, wide bandwidths, and hardware impairments~\cite{smida_fd_phy_2024,roberts_wcm}.

To address these challenges, recent work such as \cite{huang_LearningBasedHybrid_2021a,roberts_lonestar, hernangomez_CISSIRBeam_2025, bilbao_DeepUnfoldingPowered_2024,lopez_analog_2022} has explored the use of beamforming to cancel \SI, exploiting the high spatial degrees of freedom afforded by \mmwave and massive \MIMO transceivers.
While these techniques show promise, most rely on an accurate estimate of the \SI channel to design transmit and receive beams suitable for \fd operation.
This can be problematic in real-world systems, since the dimensionality of the \MIMO \SI channel scales quadratically with the number of antennas at the \BS, which is typically on the order of tens or hundreds in the case of \mmwave and massive \MIMO.
Barring sufficiently static channels, explicit estimation of the \SI channel would therefore consume substantial pilot overhead, likely rendering it impractical in 5G/6G networks.

To our knowledge, the extent of existing research on \fd beam design without explicit \SI channel knowledge is limited to \cite{roberts_steer, roberts_steerp, kong_active_fd}.
The schemes in \cite{roberts_steer, roberts_steerp} employ an exhaustive search over a small spatial grid, which can involve 50--500+ measurements per user pair, thus still incurring a significant overhead.
In \cite{kong_active_fd}, a \gls{lstm} network is used to sequentially design a set of user-specific probing beams for implicit estimation of \SI and of the downlink and uplink channels.
The primary shortcoming of this approach is its poor scaling in multi-user scenarios, since the user-specific probing procedure necessitates additional feedback/coordination.

Taking inspiration from \cite{heng_GridFreeMIMO_2024, yu_rl_2022}, where site-specific codebooks are designed to reduce beam alignment overhead in \hd systems, we propose a novel beam learning framework that bypasses explicit \SI channel estimation when designing beams to serve \DL and \UL in a \fd fashion.
Our approach jointly optimizes a pair of transmit and receive probing codebooks for implicit \SI channel estimation and a deep learning model that synthesizes the final serving beams.
Notably, the user-agnostic nature of these probing beams allows the same implicit \SI channel knowledge to be reused across many users,
thereby minimizing the overhead of our solution.
Simulations with ray-traced channel models demonstrate that our approach can design transmit and receive beams which approach the \fd capacity, while requiring only 3--25\% of the measurements needed for explicit \SI channel estimation.

%% file: sec-system-model.tex
\section{System Model}
\label{sec:system-model}

\begin{figure}[htbp]
  \centering
  \includegraphics[width=0.75\columnwidth]{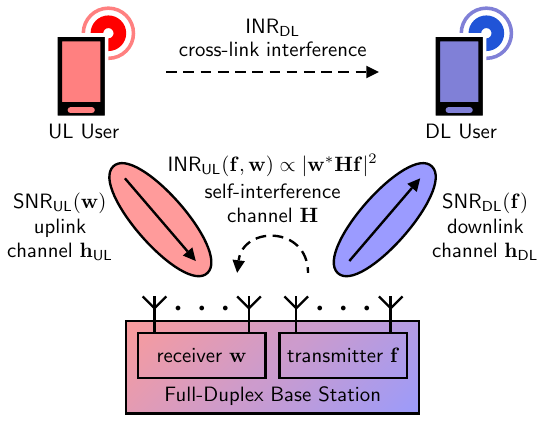}
  \caption{An in-band \fd \BS transmits to a \DL user while receiving from an \UL user across the same frequency band. This work aims to optimally design $\vf$ and $\vw$ without explicitly estimating $\mH$.}
  \label{fig:system-model}
\end{figure}

In this work, we consider a \BS that aims to concurrently transmit to a \DL user while receiving from an \UL user on the same frequency resources.
The \BS is equipped with separate antenna arrays for transmission and reception, containing $\Nt$ and $\Nr$ antennas, respectively.
Each antenna array is driven by a single \rf chain and analog beamforming weights to dynamically steer beams, realized in hardware using an analog phase shifter and a variable attenuator at each antenna.
Let us denote by $\vf \in \setvectorcomplex{\Nt}$ and $\vw \in \setvectorcomplex{\Nr}$ the transmit and receive beamforming weight vectors applied by the \BS, each subject to a per-antenna power constraint as
\begin{align}
  \label{eq:norm-f-2}
  |\vf_i| & \leq 1 \ \forall \ i \in \{1, \ldots, \Nt\},
  \\
  \label{eq:norm-w-2}
  |\vw_j| & \leq 1 \ \forall \ j  \in \{1, \ldots, \Nr\}.
\end{align}
Users are assumed to have a single omnidirectional antenna for simplicity, but this work readily extends to multi-antenna users.
In serving any given \DL user, the \BS transmits with power $P_{\dl}$ using its transmit beam $\vf$ across a channel $\vh_\dl \in \complex^{\Nt \times 1}$.
Simultaneously, the \BS receives from an \UL user, which transmits with power $P_{\ul}$, across a channel $\vh_\ul \in \complex^{\Nr \times 1}$ using its receive beam $\vw$.
With this, we can express the \DL and \UL \gpsnr as
\begin{align}
  \label{eq:snr-dl}
  \snr_\dl \qty(\vf) & = \frac{P_{\dl} \, | \vh_\dl^{*}\vf |^{2}}{\Nt \, \pnoise_{\dl}},
  \\
  \label{eq:snr-ul}
  \snr_\ul \qty(\vw) & = \frac{P_{\ul} \, | \vw^{*}\vh_\ul |^{2}}{\normtwo{\vw}^{2} \, \pnoise_{\ul}},
\end{align}
where $\pnoise_{\dl}$ and $\pnoise_\ul$ denote the per-antenna additive noise powers at each antenna on the \DL and \UL, respectively.

While operating in a \fd fashion, \SI couples from the transmit array of the \BS to its receive array across the \MIMO \SI channel $\mH \in \complex^{\Nr \times \Nt}$.
The \SI channel $\mH$ follows a Rician channel model as the sum of a \LOS component $\mH_\los$ and a \NLOS component $\mH_{\nlos}$, i.e.,
\begin{align} \label{eq:rician-channel}
  \mH = \sqrt{\frac{\kappa}{\kappa + 1}} \, \mH_{\los} + \sqrt{\frac{1}{\kappa + 1}} \, \mH_{\nlos}.
\end{align}
The \LOS component is meant to capture near-field coupling between the \BS's arrays, whereas the \NLOS component represents reflections off nearby surroundings.
While both components may fluctuate with time, the \NLOS component is likely to vary more quickly according to the dynamics of the environment in which the \BS is deployed.

To quantify the severity of \SI, we normalize its power by the noise floor, defining the \INR at the \BS receiver as
\begin{equation}
  \label{eq:inr}
  \inr_\ul \qty(\vf, \vw) = \frac{P_\dl \, |\vw^{*}\mH
  \vf |^{2}}{\Nt \, \normtwo{\vw}^{2} \, \pnoise_{\ul}}.
\end{equation}
Similar to $\inr_\ul$, the \DL user experiences cross-link interference from the \UL user's transmission, expressed as
\begin{equation}
  \label{eq:inr-dl}
  \inr_\dl = \frac{P_{\ul} \, |h|^{2}}{\pnoise_{\dl}},
\end{equation}
where $h \in \complex$ is the channel from the \UL user to the \DL user.
Treating \SI as noise, the \glspl{sinr} on the \DL and \UL are
\begin{align}
  \label{eq:sinr-dl}
  \sinr_\dl \qty(\vf)      & = \frac{\snr_\dl \qty(\vf)}{1 + \inr_\dl},
  \\
  \label{eq:sinr-ul}
  \sinr_\ul \qty(\vf, \vw) & = \frac{\snr_\ul \qty(\vw)}{1 + \inr_\ul \qty(\vf, \vw)}.
\end{align}
The achievable \DL and \UL spectral efficiencies are correspondingly
\begin{align}
  \label{eq:se-dl}
  R_\dl(\vf)     & = \logtwo{1+\sinr_\dl \qty(\vf)},
  \\
  \label{eq:se-ul}
  R_\ul(\vf,\vw) & = \logtwo{1+\sinr_\ul \qty(\vf, \vw)}.
\end{align}
As a measure of overall performance of the \fd system, we sum the \DL and \UL spectral efficiencies as
\begin{equation}
  \label{eq:sse}
  R(\vf, \vw) = R_\dl(\vf) + R_\ul(\vf, \vw),
\end{equation}
which we seek to maximize in the next section through the design of $\vf$ and $\vw$.

%% file: sec-contribution.tex
\section{Full-Duplex Beam Learning} \label{sec:contribution}

\begin{figure*}[htbp]
  \centering
  \includegraphics[width=\linewidth,height=\textheight,keepaspectratio]{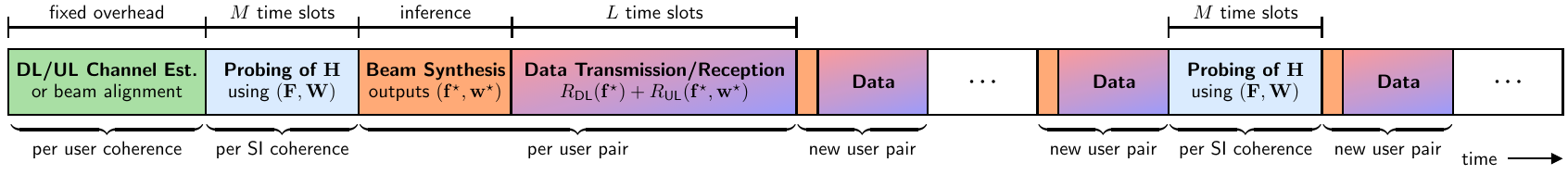}
  \caption{Timeline of the envisioned use of our proposed scheme, with one time slot defined as the time to collect a single probing measurement across $\mH$.}
  \label{fig:timeline}
\end{figure*}

In this section, we present a novel approach to designing the transmit and receive beams $(\vf,\vw)$ of the full-duplex \BS in pursuit of maximizing \sse.
A key distinction of our approach is in its use of deep learning to design these beams without explicitly estimating the self-interference channel $\mH \in \setmatrixcomplex{\Nr}{\Nt}$.
This is in contrast to most prior work such as \cite{huang_LearningBasedHybrid_2021a,roberts_lonestar,hernangomez_CISSIRBeam_2025,prelcic_2019_hybrid}, which require an accurate estimate of $\mH$ to design $\vf$ and $\vw$.
Such estimation can consume substantial pilot overhead, especially when $\mH$ changes frequently, e.g., due to dynamics in the environment.

Our proposed approach addresses this challenge by training a neural network $\sD(\cdot)$ to synthesize near-optimal transmit and receive beams $(\vf\opt,\vw\opt)$ using \textit{implicit} knowledge of the self-interference channel.
To optimize acquisition of such implicit channel knowledge, we jointly optimize the beam synthesizer $\sD(\cdot)$ and a set of $M$ pairs of transmit and receive beams $\braces{(\vf_m,\vw_m)}_{m=1}^M$.
These $M$ pairs of probing beams collect $M \ll \Nt\Nr$ scalar measurements of self-interference across $\mH$, which are then fed into the beam synthesizer, along with downlink/uplink channel knowledge, which we assume is acquired from conventional beam alignment or other channel state information mechanisms.

\subsection{Envisioned Solution}
The envisioned use of our proposed approach may take a variety of forms, one of which is illustrated in \figref{fig:timeline}.
In this scenario, the \SI channel $\mH$ fades on a time scale faster than the user channels.
As a result, \DL and \UL channel knowledge can be acquired relatively infrequently compared to probing of $\mH$.
In cases where $\mH$ is sufficiently stable, however, a single set of $M$ probing measurements may be reused across a subset of user pairs, thereby saving on overhead.

\textbf{Downlink and uplink channel acquisition:}
Our proposed framework begins by utilizing existing mechanisms for obtaining \DL and \UL channel knowledge, such as beam alignment in 5G networks \cite{ethan_beam}.
In general, this channel knowledge can be either explicit estimates of $\vh_\dl$ and $\vh_\ul$ or implicit knowledge of some form, such as beam alignment measurements.
As such, let us refer to it as simply $\vy_\dl$ and $\vy_\ul$.
Note that our approach will take as input downlink and uplink channel knowledge for a \textit{single} user pair, but this channel knowledge may be collected beforehand for \textit{all} user pairs the base station wishes to serve over some time horizon of interest.
Naturally, this step would need to be repeated commensurate with the coherence time of the users' channels.

\textbf{Probing of self-interference:}
The next component of our proposed scheme involves the \BS collecting implicit knowledge of the \SI channel $\mH$ through a series of $M$ probing measurements.
To collect the $m$-th probing measurement $z_m \in \setcomplex$, suppose the \BS transmits with beam $\vf_m$ and receives using beam $\vw_m$.
Then, the received \SI can be expressed as
\begin{equation}
  \label{eq:si-probing}
  z_m = \sqrt{\frac{P_\dl}{\Nt}} \, \vw_m^* \mH \vf_m + \vw_m^* \vn_m,
\end{equation}
where $\vn_m \sim \distcgauss{\mat{0}}{\sigma^2_\ul \mI}$ denotes the complex Gaussian noise vector at the receive array during the $m$-th measurement.

Upon stacking the probing beams into matrices $\mF$ and $\mW$
\begin{align}
  \label{eq:codebook-F}
  \mF & = \qty[ \vf_1 \ \cdots \ \vf_M ] \in \complex^{\Nt \times M}
  \\
  \label{eq:codebook-W}
  \mW & = \qty[ \vw_1 \ \cdots \ \vw_M ] \in \complex^{\Nr \times M},
\end{align}
the received signal vector $\vz = \qty[z_1 \, \cdots \, z_M]\trans \in \complex^{M \times 1}$ can be expressed compactly as
\begin{align}
  \label{eq:rec-sig-vec}
  \vz & = \sqrt{\frac{P_\dl}{\Nt}} \, \diag{\mW\ctrans \mH \mF}
  + \diag{\mW\ctrans \mN},
\end{align}
with $\mN \in \setmatrixcomplex{\Nr}{M}$ a matrix whose $m$-th column is $\vn_m$.
As will be clear shortly, the implicit channel knowledge contained in $\vz$ will be used to design the transmit and receive beams that are used by the \BS to serve the \DL and \UL users.
Since explicit estimation of $\mH$ would in general require $\Nt\Nr$ measurements, we will seek $M \ll \Nt\Nr$ and, in this pursuit, will optimize $\mF$ and $\mW$ to extract information that is most useful in designing $\vf$ and $\vw$.

\textbf{Beam synthesis:}
With implicit knowledge of $\mH$ in hand, along with downlink and uplink channel knowledge, the last stage of our proposed approach is to synthesize the final serving beams.
This is accomplished through a function $\sD(\cdot)$, which takes as input $\vz$, $\vy_\dl$, and $\vy_\ul$ and outputs $(\vf\opt,\vw\opt) = \sD\parens{\vz,\vy_\dl,\vy_\ul}$.

\subsection{Problem Formulation} \label{sec:problem}

Assembling together the components laid forth, our envisioned solution can be found by solving the following problem.%
\begin{subequations} \label{eq:problem}
  \begin{align}
    \label{eq:objective}
    \max_{\mF, \mW, \mathcal{D}(\cdot)}
    \quad
     & \expect_{\mH, \vh_\dl, \vh_\ul} \brackets{R_\dl(\vf \opt) + R_\ul(\vf \opt, \vw \opt)}
    \\ \subjectto \quad
     & \, \parens{\vf \opt, \vw \opt} = \mathcal{D}(\vz, \vy_\dl, \vy_\ul)
    \label{eq:constr-f}
    \\ \quad
     & \norm{\vf \opt}_{\max}, \norm{\vw \opt}_{\max} \leq 1
    \label{eq:constr-F}
    \\ \quad & \norm{\mF}_{\max}, \norm{\mW}_{\max} \leq 1
  \end{align}
\end{subequations}
Here, $\norm{\cdot}_{\max} = \max_{i,j}{|[\cdot]_{i,j}|}$ denotes the max norm.
Constraints \eqref{eq:constr-f} and \eqref{eq:constr-F} ensure that the synthesized serving beams $(\vf \opt, \vw \opt)$ and the probing codebooks $(\mF, \mW)$ adhere to the per-antenna power constraints of \eqref{eq:norm-f-2} and \eqref{eq:norm-w-2}.
By maximizing the expectation of \sse, we aim to find a single design of $\mF$, $\mW$, and $\sD(\cdot)$ that generalizes across user pairs and channel realizations.
Solving problem \eqref{eq:problem} within the scope of an individual \BS yields a \textit{site-specific} solution which exploits the channel conditions at that particular site within the network.

\subsection{Joint Probing Codebook Design and Beam Synthesis} \label{sec:ml-model}

\begin{figure*}[htp]
  \centering
  \scriptsize
  \fontfamily{cmss}\selectfont
  \includegraphics{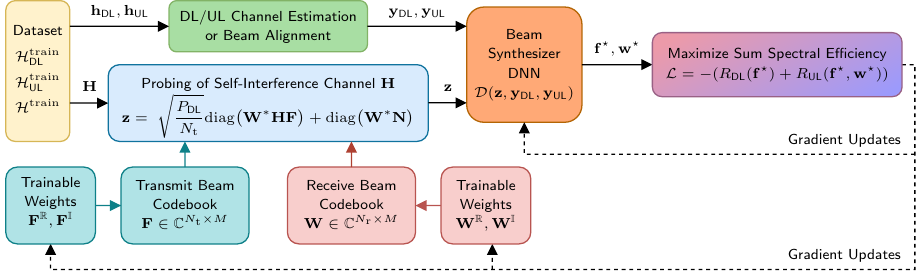}
  \caption{Proposed end-to-end \NN model for \fd beam design.}
  \label{fig:ml-model}
\end{figure*}

To solve problem \eqref{eq:problem} and realize the envisioned solution, we propose a deep learning-based framework that jointly optimizes the probing codebooks $(\mF,\mW)$ and the beam synthesizer $\mathcal{D}(\cdot)$ in an end-to-end manner.
The overall architecture, depicted in Fig.~\ref{fig:ml-model}, consists of trainable probing codebooks and a multi-layer neural network that parameterizes $\mathcal{D}(\cdot)$.
Through unsupervised training, the model tailors $\mF$ and $\mW$ to gather channel information that is most useful to the beam synthesizer in its design of $\vf\opt$ and $\vw\opt$ to maximize the \sse.
By training over the channel distributions of a single site, $\mF$, $\mW$, and $\sD(\cdot)$ are all optimized to the particular \BS at which they are deployed.

Recall, for $\mF$ and $\mW$ to be realizable in physical systems, they must be constrained to satisfy the per-antenna element magnitude limits given in \eqref{eq:constr-F}.
However, directly optimizing the complex entries of $\mF$ and $\mW$ via gradient descent does not guarantee these constraints.
Thus, we adopt an approach similar to \cite{heng_GridFreeMIMO_2024}, where the real and imaginary components of the codebooks are updated separately during training.
After each update, the codebooks are projected element-wise to satisfy \eqref{eq:constr-F}, i.e.,
\begin{equation}
  \label{eq:codebook-train}
  [\mF]_{i,j} = \frac{[{\mF^\real + \j \mF^\imaginary}]_{i,j}}{\max (1, | [{\mF^\real + \j \mF^\imaginary}]_{i,j} |)},
\end{equation}
where $\mF^\real$ and $\mF^\imaginary$ are the trainable weights that represent the real and imaginary parts of $\mF$, respectively.
$\mW$ is constructed similarly.
We use this same projection to ensure that the serving beams $\vf\opt$ and $\vw\opt$ are physically realizable.

The entire model, encompassing the learnable codebooks $(\mF,\mW)$ and the beam synthesizer network $\sD(\cdot)$, is trained end-to-end on a dataset $(\mathcal{H}^\mathrm{train}, \mathcal{H_\dl^\mathrm{train}}, \mathcal{H_\ul^\mathrm{train}})$ representative of the target operational environment.
The objective is to maximize the expectation of the \sse defined in \eqref{eq:sse}, which corresponds to minimizing the loss function
\begin{equation}
  \label{eq:loss}
  \begin{aligned}
    \mathcal{L}
     & = -\mathbb{E} \qty[R_\dl(\vf \opt) + R_\ul(\vf \opt, \vw \opt)].
  \end{aligned}
\end{equation}
After training, the optimized codebooks $\mF$ and $\mW$ can be directly used by the \BS to collect probing measurements $\vz$ according to \eqref{eq:si-probing}.
The trained beam synthesizer then uses these measurements to generate serving beams $(\vf\opt, \vw\opt)$ for any given user pair.
Once deployed, the only computational cost associated with our solution is that of beam synthesis $\sD(\cdot)$, which amounts to inference through a neural network.

%% file: sec-numerical-results.tex
\section{Numerical Results} \label{sec:numerical-results}

\begin{figure}[t]
  \centering
  {\includegraphics[width=\linewidth]{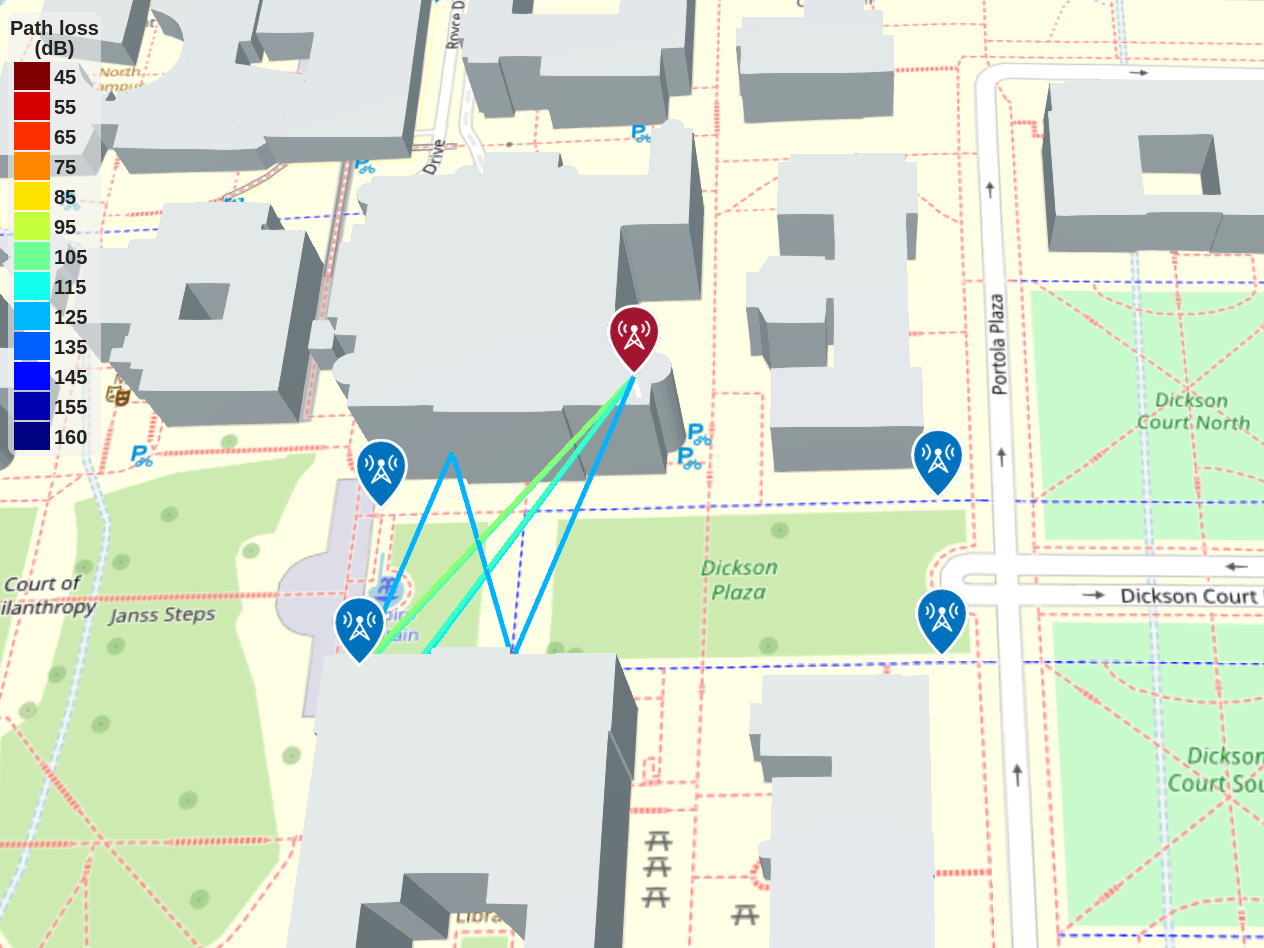}}
  \caption{Simulation scenario with a \BS and multiple users. Users are arranged in an equidistant grid on the ground bounded by the blue markers, with the \BS placed on a rooftop at the red marker.
    The traced rays for one user are visualized.}
  \label{fig:scenario}
\end{figure}

To evaluate our proposed method, we consider a 28~GHz \BS whose \TX and \RX are each equipped with a 4$\times$4 half-wavelength \acrlong{upa} and placed side by side, with their array centers separated by $10$ wavelengths horizontally. 
To simulate the downlink and uplink channels, we use ray-tracing software~\cite{matlab_comm_toolbox_2024b} to deploy users around Dickson Plaza at UCLA and mount the \BS atop Royce Hall, as illustrated in \figref{fig:scenario}.
Users are distributed within the field-of-view spanning $-60^\circ$ to $60^\circ$ of the \BS, and the \BS is down-tilted towards the center of the user distribution such that the median elevation is $0^\circ$.

The \SI channel is modeled according to \eqref{eq:rician-channel}, where the \LOS component $\mH_\los$ is modeled as a spherical-wave channel \cite{spherical_2005}, and the \NLOS component $\mH_\nlos$ is comprised of a number of reflections off the environment.
We consider both a finite number of reflections (each with amplitude drawn from a standard complex Gaussian) and an infinite number (i.e., Rayleigh fading).
Since there are 16 antennas at the transmitter and receiver of the \BS, the \SI channel is of size 16$\times$16 and therefore would in general require 256 scalar measurements to explicitly estimate.

At any given time, the \BS aims to serve a pair of \DL and \UL users in a full-duplex fashion.
While this is a predominantly \LOS environment, there exist multiple \NLOS paths to each user.
We assume the \BS conducts beam alignment to obtain knowledge of the \LOS path (i.e., angle of departure/arrival) to each user; in this case, $\vy_\dl$ and $\vy_\ul$ are the \LOS channel vectors to the users. 
The ignored \NLOS components thus represent downlink/uplink channel estimation error.
To normalize the performance evaluation, we scale ${\vh_\dl}$, ${\vh_\ul}$, and ${\mH}$ so that the maximum \DL and \UL \SNR specified in \eqref{eq:snr-dl} and \eqref{eq:snr-ul} are equal to 10~dB and the maximum \INR in \eqref{eq:inr} is 40~dB.
We ignore cross-link interference $\inr_\dl$ to focus solely on the effects of self-interference at the \BS.

The probing codebooks are implemented according to \eqref{eq:codebook-train},
and the beam synthesizer is parameterized by a fully connected neural network that takes the concatenated vector $\qty(\vz, \vy_\dl, \vy_\ul)$ as input, with 4 hidden layers and the number of neurons in each layer scaled based on $\Nt$ and $\Nr$, as shown in \tabref{tab:parameters}.
The model is trained end-to-end with a batch size of 4096 using \texttt{Adam} optimizer.
A learning rate of 0.001 is used for the beam synthesizer, and a learning rate of 0.0015 is used for the codebooks in conjunction with a cosine annealing scheduler with a period of 5000 batches and a minimum learning rate of 0.0005.
The model is trained until convergence before being tested against 20,000 unseen channel realizations to evaluate its performance.
The default parameters used in simulations are summarized in \tabref{tab:parameters}, unless specified otherwise.

\begin{table}[t]
  \centering
  \caption{Default simulation and implementation parameters}
  \label{tab:parameters}
  \renewcommand{\arraystretch}{0.8}
  \begin{tabular}{m{.53\linewidth} m{.37\linewidth}}
    \toprule
    LOS component, $\mH_\los$   & Spherical-wave channel \cite{spherical_2005}
    \\ \midrule
    NLOS component, $\mH_\nlos$ & Sum of 64 reflections
    \\ \midrule
    \Acrlong{si} Rician factor, $\kappa$
                                & 0 dB
    \\ \midrule
    Maximum \DL and \UL \SNR    & 10 dB
    \\ \midrule
    Maximum INR                 & 40 dB
    \\ \midrule
    Number of neurons in each layer of $\mathcal{D}(\cdot)$
                                & $\qty[16, 16, 8, 8] \times (\Nt + \Nr)$
    \\ \bottomrule
  \end{tabular}
\end{table}

We compare the performance of our model against
(i) {\gls{mrt} and \gls{mrc} using estimated \DL and \UL channels}, and
(ii) {the \fd capacity}, taken as the sum of the \DL and \UL capacity, which can only be achieved by \gls{mrt}+\gls{mrc} with perfect \DL and \UL channel knowledge in the absence of \SI.

\begin{figure}[t]
  \centering
  {\input{figs/Kappa_Mt.pgf}}
  \caption{Sum spectral efficiency as a function of the self-interference channel Rician factor $\kappa$ for various numbers of probing measurements $M$, where the number of reflections in $\mH_\nlos$ is 64 and $\mH_\los$ is the spherical-wave model.}
  \label{fig:kappa}
\end{figure}
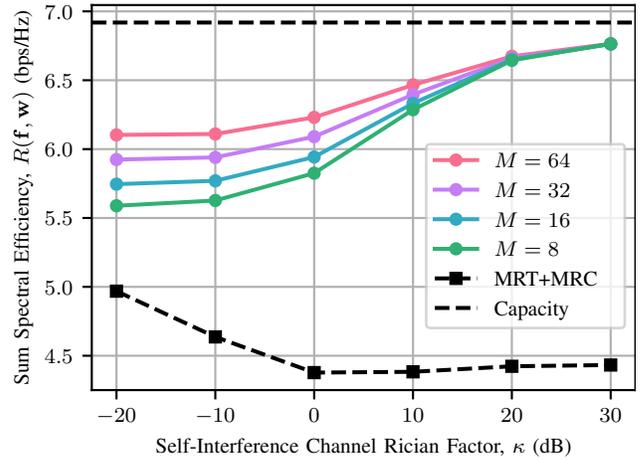

First, we investigate the impact of the \SI channel conditions on the \sse by varying the Rician factor $\kappa$, as shown in \figref{fig:kappa}.
With only 8--64 measurements of \SI, we can see appreciable performance across $\kappa$, which confirms that our approach is indeed able to perform well with a fraction of the measurements needed for explicit estimation of $\mH$.
We can also observe that our model provides improved \sse as $\kappa$ increases.
This cannot be attributed to favorable \SI conditions, as the MRT+MRC baseline trends in the opposite direction.
Instead, as the \SI channel becomes dominated by the \LOS component under high $\kappa$, the channel becomes more deterministic.
As a result, the model is able to infer a more accurate channel representation, leading to improved performance.
Conversely, in the low-$\kappa$ regime, where the \SI channel becomes more stochastic, additional probing measurements yield more pronounced gains in \sse.

\begin{figure}[t]
  \centering
  {\input{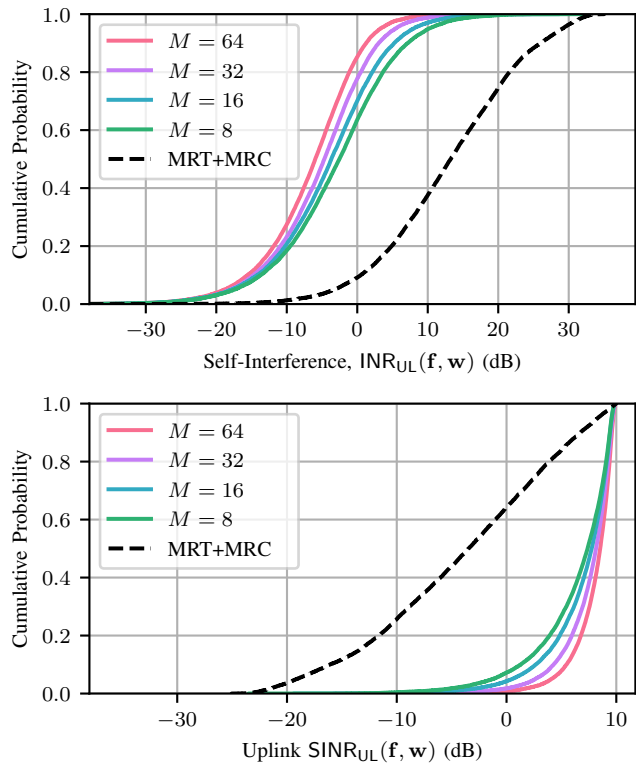}}
  \caption{CDFs of (a) self-interference $\inr_\ul$ and (b) uplink $\sinr_\ul$ for different numbers of probing measurements $M$, where $\kappa = 0$~dB.}
  \label{fig:cdf}
\end{figure}

Further evaluating the impact of $M$ on the model's performance, \figref{fig:cdf} displays the \glspl{cdf} of $\inr_\ul$ and $\sinr_\ul$ for different numbers of probing measurements $M$. 
The MRT+MRC baseline incurs approximately 15 dB of \INR, which severely degrades \SINR and renders \fd operation infeasible.
In contrast, our model achieves an \INR reduction of nearly 20 dB with only $M=16$ probing beams, reducing self-interference to below the noise floor the majority of the time and increasing \UL \SINR significantly---a trend that continues with more measurements. 
This demonstrates our model's ability to effectively mitigate \SI with a limited number of probing beams, offering a favorable trade-off between performance and measurement overhead.

\begin{figure}[t]
  \centering
  {\input{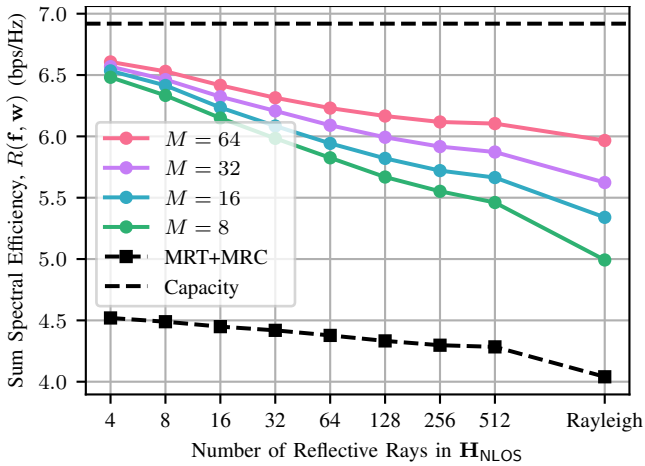}}
  \caption{Sum spectral efficiency as a function of the number of reflective rays in the \NLOS component of the \SI channel for different numbers of probing beams $M$, where $\kappa = 0$~dB.}
  \label{fig:nrays}
\end{figure}

In \figref{fig:nrays}, we train and test our model across different \SI channel configurations by varying the number of reflections in the \NLOS component $\mH_\nlos$.
This is particularly relevant to consider because only the \NLOS portion of \SI meaningfully changes with time, making the estimation of this component the principal role of probing.
With fewer rays, fewer probing measurements $M$ are needed to attain high \sse and increasing the number of probing measurements offers little gain.
As the number of rays increases, the benefit of more measurements appears, especially when an infinite number of rays are present in the case of Rayleigh fading.
However, in such regimes, the self-interference channel is particularly rich, making it difficult to reliably obtain enough implicit knowledge from only $M=64$ measurements, considering the channel matrix is full rank of size $16 \times 16$.

%% file: figs/kappa_Mt.pgf
\begingroup%
\makeatletter%
\begin{pgfpicture}%
\pgfpathrectangle{\pgfpointorigin}{\pgfqpoint{3.260000in}{2.385070in}}%
\pgfusepath{use as bounding box, clip}%
\begin{pgfscope}%
\pgfsetbuttcap%
\pgfsetmiterjoin%
\definecolor{currentfill}{rgb}{1.000000,1.000000,1.000000}%
\pgfsetfillcolor{currentfill}%
\pgfsetlinewidth{0.000000pt}%
\definecolor{currentstroke}{rgb}{1.000000,1.000000,1.000000}%
\pgfsetstrokecolor{currentstroke}%
\pgfsetdash{}{0pt}%
\pgfpathmoveto{\pgfqpoint{0.000000in}{0.000000in}}%
\pgfpathlineto{\pgfqpoint{3.260000in}{0.000000in}}%
\pgfpathlineto{\pgfqpoint{3.260000in}{2.385070in}}%
\pgfpathlineto{\pgfqpoint{0.000000in}{2.385070in}}%
\pgfpathlineto{\pgfqpoint{0.000000in}{0.000000in}}%
\pgfpathclose%
\pgfusepath{fill}%
\end{pgfscope}%
\begin{pgfscope}%
\pgfsetbuttcap%
\pgfsetmiterjoin%
\definecolor{currentfill}{rgb}{1.000000,1.000000,1.000000}%
\pgfsetfillcolor{currentfill}%
\pgfsetlinewidth{0.000000pt}%
\definecolor{currentstroke}{rgb}{0.000000,0.000000,0.000000}%
\pgfsetstrokecolor{currentstroke}%
\pgfsetstrokeopacity{0.000000}%
\pgfsetdash{}{0pt}%
\pgfpathmoveto{\pgfqpoint{0.414740in}{0.362654in}}%
\pgfpathlineto{\pgfqpoint{3.260000in}{0.362654in}}%
\pgfpathlineto{\pgfqpoint{3.260000in}{2.379630in}}%
\pgfpathlineto{\pgfqpoint{0.414740in}{2.379630in}}%
\pgfpathlineto{\pgfqpoint{0.414740in}{0.362654in}}%
\pgfpathclose%
\pgfusepath{fill}%
\end{pgfscope}%
\begin{pgfscope}%
\pgfpathrectangle{\pgfqpoint{0.414740in}{0.362654in}}{\pgfqpoint{2.845260in}{2.016976in}}%
\pgfusepath{clip}%
\pgfsetrectcap%
\pgfsetroundjoin%
\pgfsetlinewidth{0.803000pt}%
\definecolor{currentstroke}{rgb}{0.690196,0.690196,0.690196}%
\pgfsetstrokecolor{currentstroke}%
\pgfsetdash{}{0pt}%
\pgfpathmoveto{\pgfqpoint{0.544070in}{0.362654in}}%
\pgfpathlineto{\pgfqpoint{0.544070in}{2.379630in}}%
\pgfusepath{stroke}%
\end{pgfscope}%
\begin{pgfscope}%
\pgfsetbuttcap%
\pgfsetroundjoin%
\definecolor{currentfill}{rgb}{0.000000,0.000000,0.000000}%
\pgfsetfillcolor{currentfill}%
\pgfsetlinewidth{0.803000pt}%
\definecolor{currentstroke}{rgb}{0.000000,0.000000,0.000000}%
\pgfsetstrokecolor{currentstroke}%
\pgfsetdash{}{0pt}%
\pgfsys@defobject{currentmarker}{\pgfqpoint{0.000000in}{-0.048611in}}{\pgfqpoint{0.000000in}{0.000000in}}{%
\pgfpathmoveto{\pgfqpoint{0.000000in}{0.000000in}}%
\pgfpathlineto{\pgfqpoint{0.000000in}{-0.048611in}}%
\pgfusepath{stroke,fill}%
}%
\begin{pgfscope}%
\pgfsys@transformshift{0.544070in}{0.362654in}%
\pgfsys@useobject{currentmarker}{}%
\end{pgfscope}%
\end{pgfscope}%
\begin{pgfscope}%
\definecolor{textcolor}{rgb}{0.000000,0.000000,0.000000}%
\pgfsetstrokecolor{textcolor}%
\pgfsetfillcolor{textcolor}%
\pgftext[x=0.544070in,y=0.265432in,,top]{\color{textcolor}{\rmfamily\fontsize{8.000000}{9.600000}\selectfont\catcode`\^=\active\def^{\ifmmode\sp\else\^{}\fi}\catcode`\%=\active\def%{\%}$\mathdefault{\ensuremath{-}20}$}}%
\end{pgfscope}%
\begin{pgfscope}%
\pgfpathrectangle{\pgfqpoint{0.414740in}{0.362654in}}{\pgfqpoint{2.845260in}{2.016976in}}%
\pgfusepath{clip}%
\pgfsetrectcap%
\pgfsetroundjoin%
\pgfsetlinewidth{0.803000pt}%
\definecolor{currentstroke}{rgb}{0.690196,0.690196,0.690196}%
\pgfsetstrokecolor{currentstroke}%
\pgfsetdash{}{0pt}%
\pgfpathmoveto{\pgfqpoint{1.061390in}{0.362654in}}%
\pgfpathlineto{\pgfqpoint{1.061390in}{2.379630in}}%
\pgfusepath{stroke}%
\end{pgfscope}%
\begin{pgfscope}%
\pgfsetbuttcap%
\pgfsetroundjoin%
\definecolor{currentfill}{rgb}{0.000000,0.000000,0.000000}%
\pgfsetfillcolor{currentfill}%
\pgfsetlinewidth{0.803000pt}%
\definecolor{currentstroke}{rgb}{0.000000,0.000000,0.000000}%
\pgfsetstrokecolor{currentstroke}%
\pgfsetdash{}{0pt}%
\pgfsys@defobject{currentmarker}{\pgfqpoint{0.000000in}{-0.048611in}}{\pgfqpoint{0.000000in}{0.000000in}}{%
\pgfpathmoveto{\pgfqpoint{0.000000in}{0.000000in}}%
\pgfpathlineto{\pgfqpoint{0.000000in}{-0.048611in}}%
\pgfusepath{stroke,fill}%
}%
\begin{pgfscope}%
\pgfsys@transformshift{1.061390in}{0.362654in}%
\pgfsys@useobject{currentmarker}{}%
\end{pgfscope}%
\end{pgfscope}%
\begin{pgfscope}%
\definecolor{textcolor}{rgb}{0.000000,0.000000,0.000000}%
\pgfsetstrokecolor{textcolor}%
\pgfsetfillcolor{textcolor}%
\pgftext[x=1.061390in,y=0.265432in,,top]{\color{textcolor}{\rmfamily\fontsize{8.000000}{9.600000}\selectfont\catcode`\^=\active\def^{\ifmmode\sp\else\^{}\fi}\catcode`\%=\active\def%{\%}$\mathdefault{\ensuremath{-}10}$}}%
\end{pgfscope}%
\begin{pgfscope}%
\pgfpathrectangle{\pgfqpoint{0.414740in}{0.362654in}}{\pgfqpoint{2.845260in}{2.016976in}}%
\pgfusepath{clip}%
\pgfsetrectcap%
\pgfsetroundjoin%
\pgfsetlinewidth{0.803000pt}%
\definecolor{currentstroke}{rgb}{0.690196,0.690196,0.690196}%
\pgfsetstrokecolor{currentstroke}%
\pgfsetdash{}{0pt}%
\pgfpathmoveto{\pgfqpoint{1.578710in}{0.362654in}}%
\pgfpathlineto{\pgfqpoint{1.578710in}{2.379630in}}%
\pgfusepath{stroke}%
\end{pgfscope}%
\begin{pgfscope}%
\pgfsetbuttcap%
\pgfsetroundjoin%
\definecolor{currentfill}{rgb}{0.000000,0.000000,0.000000}%
\pgfsetfillcolor{currentfill}%
\pgfsetlinewidth{0.803000pt}%
\definecolor{currentstroke}{rgb}{0.000000,0.000000,0.000000}%
\pgfsetstrokecolor{currentstroke}%
\pgfsetdash{}{0pt}%
\pgfsys@defobject{currentmarker}{\pgfqpoint{0.000000in}{-0.048611in}}{\pgfqpoint{0.000000in}{0.000000in}}{%
\pgfpathmoveto{\pgfqpoint{0.000000in}{0.000000in}}%
\pgfpathlineto{\pgfqpoint{0.000000in}{-0.048611in}}%
\pgfusepath{stroke,fill}%
}%
\begin{pgfscope}%
\pgfsys@transformshift{1.578710in}{0.362654in}%
\pgfsys@useobject{currentmarker}{}%
\end{pgfscope}%
\end{pgfscope}%
\begin{pgfscope}%
\definecolor{textcolor}{rgb}{0.000000,0.000000,0.000000}%
\pgfsetstrokecolor{textcolor}%
\pgfsetfillcolor{textcolor}%
\pgftext[x=1.578710in,y=0.265432in,,top]{\color{textcolor}{\rmfamily\fontsize{8.000000}{9.600000}\selectfont\catcode`\^=\active\def^{\ifmmode\sp\else\^{}\fi}\catcode`\%=\active\def%{\%}$\mathdefault{0}$}}%
\end{pgfscope}%
\begin{pgfscope}%
\pgfpathrectangle{\pgfqpoint{0.414740in}{0.362654in}}{\pgfqpoint{2.845260in}{2.016976in}}%
\pgfusepath{clip}%
\pgfsetrectcap%
\pgfsetroundjoin%
\pgfsetlinewidth{0.803000pt}%
\definecolor{currentstroke}{rgb}{0.690196,0.690196,0.690196}%
\pgfsetstrokecolor{currentstroke}%
\pgfsetdash{}{0pt}%
\pgfpathmoveto{\pgfqpoint{2.096030in}{0.362654in}}%
\pgfpathlineto{\pgfqpoint{2.096030in}{2.379630in}}%
\pgfusepath{stroke}%
\end{pgfscope}%
\begin{pgfscope}%
\pgfsetbuttcap%
\pgfsetroundjoin%
\definecolor{currentfill}{rgb}{0.000000,0.000000,0.000000}%
\pgfsetfillcolor{currentfill}%
\pgfsetlinewidth{0.803000pt}%
\definecolor{currentstroke}{rgb}{0.000000,0.000000,0.000000}%
\pgfsetstrokecolor{currentstroke}%
\pgfsetdash{}{0pt}%
\pgfsys@defobject{currentmarker}{\pgfqpoint{0.000000in}{-0.048611in}}{\pgfqpoint{0.000000in}{0.000000in}}{%
\pgfpathmoveto{\pgfqpoint{0.000000in}{0.000000in}}%
\pgfpathlineto{\pgfqpoint{0.000000in}{-0.048611in}}%
\pgfusepath{stroke,fill}%
}%
\begin{pgfscope}%
\pgfsys@transformshift{2.096030in}{0.362654in}%
\pgfsys@useobject{currentmarker}{}%
\end{pgfscope}%
\end{pgfscope}%
\begin{pgfscope}%
\definecolor{textcolor}{rgb}{0.000000,0.000000,0.000000}%
\pgfsetstrokecolor{textcolor}%
\pgfsetfillcolor{textcolor}%
\pgftext[x=2.096030in,y=0.265432in,,top]{\color{textcolor}{\rmfamily\fontsize{8.000000}{9.600000}\selectfont\catcode`\^=\active\def^{\ifmmode\sp\else\^{}\fi}\catcode`\%=\active\def%{\%}$\mathdefault{10}$}}%
\end{pgfscope}%
\begin{pgfscope}%
\pgfpathrectangle{\pgfqpoint{0.414740in}{0.362654in}}{\pgfqpoint{2.845260in}{2.016976in}}%
\pgfusepath{clip}%
\pgfsetrectcap%
\pgfsetroundjoin%
\pgfsetlinewidth{0.803000pt}%
\definecolor{currentstroke}{rgb}{0.690196,0.690196,0.690196}%
\pgfsetstrokecolor{currentstroke}%
\pgfsetdash{}{0pt}%
\pgfpathmoveto{\pgfqpoint{2.613350in}{0.362654in}}%
\pgfpathlineto{\pgfqpoint{2.613350in}{2.379630in}}%
\pgfusepath{stroke}%
\end{pgfscope}%
\begin{pgfscope}%
\pgfsetbuttcap%
\pgfsetroundjoin%
\definecolor{currentfill}{rgb}{0.000000,0.000000,0.000000}%
\pgfsetfillcolor{currentfill}%
\pgfsetlinewidth{0.803000pt}%
\definecolor{currentstroke}{rgb}{0.000000,0.000000,0.000000}%
\pgfsetstrokecolor{currentstroke}%
\pgfsetdash{}{0pt}%
\pgfsys@defobject{currentmarker}{\pgfqpoint{0.000000in}{-0.048611in}}{\pgfqpoint{0.000000in}{0.000000in}}{%
\pgfpathmoveto{\pgfqpoint{0.000000in}{0.000000in}}%
\pgfpathlineto{\pgfqpoint{0.000000in}{-0.048611in}}%
\pgfusepath{stroke,fill}%
}%
\begin{pgfscope}%
\pgfsys@transformshift{2.613350in}{0.362654in}%
\pgfsys@useobject{currentmarker}{}%
\end{pgfscope}%
\end{pgfscope}%
\begin{pgfscope}%
\definecolor{textcolor}{rgb}{0.000000,0.000000,0.000000}%
\pgfsetstrokecolor{textcolor}%
\pgfsetfillcolor{textcolor}%
\pgftext[x=2.613350in,y=0.265432in,,top]{\color{textcolor}{\rmfamily\fontsize{8.000000}{9.600000}\selectfont\catcode`\^=\active\def^{\ifmmode\sp\else\^{}\fi}\catcode`\%=\active\def%{\%}$\mathdefault{20}$}}%
\end{pgfscope}%
\begin{pgfscope}%
\pgfpathrectangle{\pgfqpoint{0.414740in}{0.362654in}}{\pgfqpoint{2.845260in}{2.016976in}}%
\pgfusepath{clip}%
\pgfsetrectcap%
\pgfsetroundjoin%
\pgfsetlinewidth{0.803000pt}%
\definecolor{currentstroke}{rgb}{0.690196,0.690196,0.690196}%
\pgfsetstrokecolor{currentstroke}%
\pgfsetdash{}{0pt}%
\pgfpathmoveto{\pgfqpoint{3.130670in}{0.362654in}}%
\pgfpathlineto{\pgfqpoint{3.130670in}{2.379630in}}%
\pgfusepath{stroke}%
\end{pgfscope}%
\begin{pgfscope}%
\pgfsetbuttcap%
\pgfsetroundjoin%
\definecolor{currentfill}{rgb}{0.000000,0.000000,0.000000}%
\pgfsetfillcolor{currentfill}%
\pgfsetlinewidth{0.803000pt}%
\definecolor{currentstroke}{rgb}{0.000000,0.000000,0.000000}%
\pgfsetstrokecolor{currentstroke}%
\pgfsetdash{}{0pt}%
\pgfsys@defobject{currentmarker}{\pgfqpoint{0.000000in}{-0.048611in}}{\pgfqpoint{0.000000in}{0.000000in}}{%
\pgfpathmoveto{\pgfqpoint{0.000000in}{0.000000in}}%
\pgfpathlineto{\pgfqpoint{0.000000in}{-0.048611in}}%
\pgfusepath{stroke,fill}%
}%
\begin{pgfscope}%
\pgfsys@transformshift{3.130670in}{0.362654in}%
\pgfsys@useobject{currentmarker}{}%
\end{pgfscope}%
\end{pgfscope}%
\begin{pgfscope}%
\definecolor{textcolor}{rgb}{0.000000,0.000000,0.000000}%
\pgfsetstrokecolor{textcolor}%
\pgfsetfillcolor{textcolor}%
\pgftext[x=3.130670in,y=0.265432in,,top]{\color{textcolor}{\rmfamily\fontsize{8.000000}{9.600000}\selectfont\catcode`\^=\active\def^{\ifmmode\sp\else\^{}\fi}\catcode`\%=\active\def%{\%}$\mathdefault{30}$}}%
\end{pgfscope}%
\begin{pgfscope}%
\definecolor{textcolor}{rgb}{0.000000,0.000000,0.000000}%
\pgfsetstrokecolor{textcolor}%
\pgfsetfillcolor{textcolor}%
\pgftext[x=1.837370in,y=0.111111in,,top]{\color{textcolor}{\rmfamily\fontsize{8.000000}{9.600000}\selectfont\catcode`\^=\active\def^{\ifmmode\sp\else\^{}\fi}\catcode`\%=\active\def%{\%}Self-Interference Channel Rician Factor, $\kappa$ (dB)}}%
\end{pgfscope}%
\begin{pgfscope}%
\pgfpathrectangle{\pgfqpoint{0.414740in}{0.362654in}}{\pgfqpoint{2.845260in}{2.016976in}}%
\pgfusepath{clip}%
\pgfsetrectcap%
\pgfsetroundjoin%
\pgfsetlinewidth{0.803000pt}%
\definecolor{currentstroke}{rgb}{0.690196,0.690196,0.690196}%
\pgfsetstrokecolor{currentstroke}%
\pgfsetdash{}{0pt}%
\pgfpathmoveto{\pgfqpoint{0.414740in}{0.542727in}}%
\pgfpathlineto{\pgfqpoint{3.260000in}{0.542727in}}%
\pgfusepath{stroke}%
\end{pgfscope}%
\begin{pgfscope}%
\pgfsetbuttcap%
\pgfsetroundjoin%
\definecolor{currentfill}{rgb}{0.000000,0.000000,0.000000}%
\pgfsetfillcolor{currentfill}%
\pgfsetlinewidth{0.803000pt}%
\definecolor{currentstroke}{rgb}{0.000000,0.000000,0.000000}%
\pgfsetstrokecolor{currentstroke}%
\pgfsetdash{}{0pt}%
\pgfsys@defobject{currentmarker}{\pgfqpoint{-0.048611in}{0.000000in}}{\pgfqpoint{-0.000000in}{0.000000in}}{%
\pgfpathmoveto{\pgfqpoint{-0.000000in}{0.000000in}}%
\pgfpathlineto{\pgfqpoint{-0.048611in}{0.000000in}}%
\pgfusepath{stroke,fill}%
}%
\begin{pgfscope}%
\pgfsys@transformshift{0.414740in}{0.542727in}%
\pgfsys@useobject{currentmarker}{}%
\end{pgfscope}%
\end{pgfscope}%
\begin{pgfscope}%
\definecolor{textcolor}{rgb}{0.000000,0.000000,0.000000}%
\pgfsetstrokecolor{textcolor}%
\pgfsetfillcolor{textcolor}%
\pgftext[x=0.166667in, y=0.504147in, left, base]{\color{textcolor}{\rmfamily\fontsize{8.000000}{9.600000}\selectfont\catcode`\^=\active\def^{\ifmmode\sp\else\^{}\fi}\catcode`\%=\active\def%{\%}$\mathdefault{4.5}$}}%
\end{pgfscope}%
\begin{pgfscope}%
\pgfpathrectangle{\pgfqpoint{0.414740in}{0.362654in}}{\pgfqpoint{2.845260in}{2.016976in}}%
\pgfusepath{clip}%
\pgfsetrectcap%
\pgfsetroundjoin%
\pgfsetlinewidth{0.803000pt}%
\definecolor{currentstroke}{rgb}{0.690196,0.690196,0.690196}%
\pgfsetstrokecolor{currentstroke}%
\pgfsetdash{}{0pt}%
\pgfpathmoveto{\pgfqpoint{0.414740in}{0.903480in}}%
\pgfpathlineto{\pgfqpoint{3.260000in}{0.903480in}}%
\pgfusepath{stroke}%
\end{pgfscope}%
\begin{pgfscope}%
\pgfsetbuttcap%
\pgfsetroundjoin%
\definecolor{currentfill}{rgb}{0.000000,0.000000,0.000000}%
\pgfsetfillcolor{currentfill}%
\pgfsetlinewidth{0.803000pt}%
\definecolor{currentstroke}{rgb}{0.000000,0.000000,0.000000}%
\pgfsetstrokecolor{currentstroke}%
\pgfsetdash{}{0pt}%
\pgfsys@defobject{currentmarker}{\pgfqpoint{-0.048611in}{0.000000in}}{\pgfqpoint{-0.000000in}{0.000000in}}{%
\pgfpathmoveto{\pgfqpoint{-0.000000in}{0.000000in}}%
\pgfpathlineto{\pgfqpoint{-0.048611in}{0.000000in}}%
\pgfusepath{stroke,fill}%
}%
\begin{pgfscope}%
\pgfsys@transformshift{0.414740in}{0.903480in}%
\pgfsys@useobject{currentmarker}{}%
\end{pgfscope}%
\end{pgfscope}%
\begin{pgfscope}%
\definecolor{textcolor}{rgb}{0.000000,0.000000,0.000000}%
\pgfsetstrokecolor{textcolor}%
\pgfsetfillcolor{textcolor}%
\pgftext[x=0.166667in, y=0.864900in, left, base]{\color{textcolor}{\rmfamily\fontsize{8.000000}{9.600000}\selectfont\catcode`\^=\active\def^{\ifmmode\sp\else\^{}\fi}\catcode`\%=\active\def%{\%}$\mathdefault{5.0}$}}%
\end{pgfscope}%
\begin{pgfscope}%
\pgfpathrectangle{\pgfqpoint{0.414740in}{0.362654in}}{\pgfqpoint{2.845260in}{2.016976in}}%
\pgfusepath{clip}%
\pgfsetrectcap%
\pgfsetroundjoin%
\pgfsetlinewidth{0.803000pt}%
\definecolor{currentstroke}{rgb}{0.690196,0.690196,0.690196}%
\pgfsetstrokecolor{currentstroke}%
\pgfsetdash{}{0pt}%
\pgfpathmoveto{\pgfqpoint{0.414740in}{1.264232in}}%
\pgfpathlineto{\pgfqpoint{3.260000in}{1.264232in}}%
\pgfusepath{stroke}%
\end{pgfscope}%
\begin{pgfscope}%
\pgfsetbuttcap%
\pgfsetroundjoin%
\definecolor{currentfill}{rgb}{0.000000,0.000000,0.000000}%
\pgfsetfillcolor{currentfill}%
\pgfsetlinewidth{0.803000pt}%
\definecolor{currentstroke}{rgb}{0.000000,0.000000,0.000000}%
\pgfsetstrokecolor{currentstroke}%
\pgfsetdash{}{0pt}%
\pgfsys@defobject{currentmarker}{\pgfqpoint{-0.048611in}{0.000000in}}{\pgfqpoint{-0.000000in}{0.000000in}}{%
\pgfpathmoveto{\pgfqpoint{-0.000000in}{0.000000in}}%
\pgfpathlineto{\pgfqpoint{-0.048611in}{0.000000in}}%
\pgfusepath{stroke,fill}%
}%
\begin{pgfscope}%
\pgfsys@transformshift{0.414740in}{1.264232in}%
\pgfsys@useobject{currentmarker}{}%
\end{pgfscope}%
\end{pgfscope}%
\begin{pgfscope}%
\definecolor{textcolor}{rgb}{0.000000,0.000000,0.000000}%
\pgfsetstrokecolor{textcolor}%
\pgfsetfillcolor{textcolor}%
\pgftext[x=0.166667in, y=1.225652in, left, base]{\color{textcolor}{\rmfamily\fontsize{8.000000}{9.600000}\selectfont\catcode`\^=\active\def^{\ifmmode\sp\else\^{}\fi}\catcode`\%=\active\def%{\%}$\mathdefault{5.5}$}}%
\end{pgfscope}%
\begin{pgfscope}%
\pgfpathrectangle{\pgfqpoint{0.414740in}{0.362654in}}{\pgfqpoint{2.845260in}{2.016976in}}%
\pgfusepath{clip}%
\pgfsetrectcap%
\pgfsetroundjoin%
\pgfsetlinewidth{0.803000pt}%
\definecolor{currentstroke}{rgb}{0.690196,0.690196,0.690196}%
\pgfsetstrokecolor{currentstroke}%
\pgfsetdash{}{0pt}%
\pgfpathmoveto{\pgfqpoint{0.414740in}{1.624985in}}%
\pgfpathlineto{\pgfqpoint{3.260000in}{1.624985in}}%
\pgfusepath{stroke}%
\end{pgfscope}%
\begin{pgfscope}%
\pgfsetbuttcap%
\pgfsetroundjoin%
\definecolor{currentfill}{rgb}{0.000000,0.000000,0.000000}%
\pgfsetfillcolor{currentfill}%
\pgfsetlinewidth{0.803000pt}%
\definecolor{currentstroke}{rgb}{0.000000,0.000000,0.000000}%
\pgfsetstrokecolor{currentstroke}%
\pgfsetdash{}{0pt}%
\pgfsys@defobject{currentmarker}{\pgfqpoint{-0.048611in}{0.000000in}}{\pgfqpoint{-0.000000in}{0.000000in}}{%
\pgfpathmoveto{\pgfqpoint{-0.000000in}{0.000000in}}%
\pgfpathlineto{\pgfqpoint{-0.048611in}{0.000000in}}%
\pgfusepath{stroke,fill}%
}%
\begin{pgfscope}%
\pgfsys@transformshift{0.414740in}{1.624985in}%
\pgfsys@useobject{currentmarker}{}%
\end{pgfscope}%
\end{pgfscope}%
\begin{pgfscope}%
\definecolor{textcolor}{rgb}{0.000000,0.000000,0.000000}%
\pgfsetstrokecolor{textcolor}%
\pgfsetfillcolor{textcolor}%
\pgftext[x=0.166667in, y=1.586405in, left, base]{\color{textcolor}{\rmfamily\fontsize{8.000000}{9.600000}\selectfont\catcode`\^=\active\def^{\ifmmode\sp\else\^{}\fi}\catcode`\%=\active\def%{\%}$\mathdefault{6.0}$}}%
\end{pgfscope}%
\begin{pgfscope}%
\pgfpathrectangle{\pgfqpoint{0.414740in}{0.362654in}}{\pgfqpoint{2.845260in}{2.016976in}}%
\pgfusepath{clip}%
\pgfsetrectcap%
\pgfsetroundjoin%
\pgfsetlinewidth{0.803000pt}%
\definecolor{currentstroke}{rgb}{0.690196,0.690196,0.690196}%
\pgfsetstrokecolor{currentstroke}%
\pgfsetdash{}{0pt}%
\pgfpathmoveto{\pgfqpoint{0.414740in}{1.985737in}}%
\pgfpathlineto{\pgfqpoint{3.260000in}{1.985737in}}%
\pgfusepath{stroke}%
\end{pgfscope}%
\begin{pgfscope}%
\pgfsetbuttcap%
\pgfsetroundjoin%
\definecolor{currentfill}{rgb}{0.000000,0.000000,0.000000}%
\pgfsetfillcolor{currentfill}%
\pgfsetlinewidth{0.803000pt}%
\definecolor{currentstroke}{rgb}{0.000000,0.000000,0.000000}%
\pgfsetstrokecolor{currentstroke}%
\pgfsetdash{}{0pt}%
\pgfsys@defobject{currentmarker}{\pgfqpoint{-0.048611in}{0.000000in}}{\pgfqpoint{-0.000000in}{0.000000in}}{%
\pgfpathmoveto{\pgfqpoint{-0.000000in}{0.000000in}}%
\pgfpathlineto{\pgfqpoint{-0.048611in}{0.000000in}}%
\pgfusepath{stroke,fill}%
}%
\begin{pgfscope}%
\pgfsys@transformshift{0.414740in}{1.985737in}%
\pgfsys@useobject{currentmarker}{}%
\end{pgfscope}%
\end{pgfscope}%
\begin{pgfscope}%
\definecolor{textcolor}{rgb}{0.000000,0.000000,0.000000}%
\pgfsetstrokecolor{textcolor}%
\pgfsetfillcolor{textcolor}%
\pgftext[x=0.166667in, y=1.947157in, left, base]{\color{textcolor}{\rmfamily\fontsize{8.000000}{9.600000}\selectfont\catcode`\^=\active\def^{\ifmmode\sp\else\^{}\fi}\catcode`\%=\active\def%{\%}$\mathdefault{6.5}$}}%
\end{pgfscope}%
\begin{pgfscope}%
\pgfpathrectangle{\pgfqpoint{0.414740in}{0.362654in}}{\pgfqpoint{2.845260in}{2.016976in}}%
\pgfusepath{clip}%
\pgfsetrectcap%
\pgfsetroundjoin%
\pgfsetlinewidth{0.803000pt}%
\definecolor{currentstroke}{rgb}{0.690196,0.690196,0.690196}%
\pgfsetstrokecolor{currentstroke}%
\pgfsetdash{}{0pt}%
\pgfpathmoveto{\pgfqpoint{0.414740in}{2.346490in}}%
\pgfpathlineto{\pgfqpoint{3.260000in}{2.346490in}}%
\pgfusepath{stroke}%
\end{pgfscope}%
\begin{pgfscope}%
\pgfsetbuttcap%
\pgfsetroundjoin%
\definecolor{currentfill}{rgb}{0.000000,0.000000,0.000000}%
\pgfsetfillcolor{currentfill}%
\pgfsetlinewidth{0.803000pt}%
\definecolor{currentstroke}{rgb}{0.000000,0.000000,0.000000}%
\pgfsetstrokecolor{currentstroke}%
\pgfsetdash{}{0pt}%
\pgfsys@defobject{currentmarker}{\pgfqpoint{-0.048611in}{0.000000in}}{\pgfqpoint{-0.000000in}{0.000000in}}{%
\pgfpathmoveto{\pgfqpoint{-0.000000in}{0.000000in}}%
\pgfpathlineto{\pgfqpoint{-0.048611in}{0.000000in}}%
\pgfusepath{stroke,fill}%
}%
\begin{pgfscope}%
\pgfsys@transformshift{0.414740in}{2.346490in}%
\pgfsys@useobject{currentmarker}{}%
\end{pgfscope}%
\end{pgfscope}%
\begin{pgfscope}%
\definecolor{textcolor}{rgb}{0.000000,0.000000,0.000000}%
\pgfsetstrokecolor{textcolor}%
\pgfsetfillcolor{textcolor}%
\pgftext[x=0.166667in, y=2.307910in, left, base]{\color{textcolor}{\rmfamily\fontsize{8.000000}{9.600000}\selectfont\catcode`\^=\active\def^{\ifmmode\sp\else\^{}\fi}\catcode`\%=\active\def%{\%}$\mathdefault{7.0}$}}%
\end{pgfscope}%
\begin{pgfscope}%
\definecolor{textcolor}{rgb}{0.000000,0.000000,0.000000}%
\pgfsetstrokecolor{textcolor}%
\pgfsetfillcolor{textcolor}%
\pgftext[x=0.111111in,y=1.371142in,,bottom,rotate=90.000000]{\color{textcolor}{\rmfamily\fontsize{8.000000}{9.600000}\selectfont\catcode`\^=\active\def^{\ifmmode\sp\else\^{}\fi}\catcode`\%=\active\def%{\%}Sum Spectral Efficiency, $R(\vf, \vw)$ (bps/Hz)}}%
\end{pgfscope}%
\begin{pgfscope}%
\pgfpathrectangle{\pgfqpoint{0.414740in}{0.362654in}}{\pgfqpoint{2.845260in}{2.016976in}}%
\pgfusepath{clip}%
\pgfsetrectcap%
\pgfsetroundjoin%
\pgfsetlinewidth{1.505625pt}%
\definecolor{currentstroke}{rgb}{0.973355,0.432701,0.565100}%
\pgfsetstrokecolor{currentstroke}%
\pgfsetdash{}{0pt}%
\pgfpathmoveto{\pgfqpoint{0.544070in}{1.699175in}}%
\pgfpathlineto{\pgfqpoint{1.061390in}{1.704246in}}%
\pgfpathlineto{\pgfqpoint{1.578710in}{1.791322in}}%
\pgfpathlineto{\pgfqpoint{2.096030in}{1.961271in}}%
\pgfpathlineto{\pgfqpoint{2.613350in}{2.111045in}}%
\pgfpathlineto{\pgfqpoint{3.130670in}{2.176772in}}%
\pgfusepath{stroke}%
\end{pgfscope}%
\begin{pgfscope}%
\pgfpathrectangle{\pgfqpoint{0.414740in}{0.362654in}}{\pgfqpoint{2.845260in}{2.016976in}}%
\pgfusepath{clip}%
\pgfsetbuttcap%
\pgfsetroundjoin%
\definecolor{currentfill}{rgb}{0.973355,0.432701,0.565100}%
\pgfsetfillcolor{currentfill}%
\pgfsetlinewidth{1.003750pt}%
\definecolor{currentstroke}{rgb}{0.973355,0.432701,0.565100}%
\pgfsetstrokecolor{currentstroke}%
\pgfsetdash{}{0pt}%
\pgfsys@defobject{currentmarker}{\pgfqpoint{-0.027778in}{-0.027778in}}{\pgfqpoint{0.027778in}{0.027778in}}{%
\pgfpathmoveto{\pgfqpoint{0.000000in}{-0.027778in}}%
\pgfpathcurveto{\pgfqpoint{0.007367in}{-0.027778in}}{\pgfqpoint{0.014433in}{-0.024851in}}{\pgfqpoint{0.019642in}{-0.019642in}}%
\pgfpathcurveto{\pgfqpoint{0.024851in}{-0.014433in}}{\pgfqpoint{0.027778in}{-0.007367in}}{\pgfqpoint{0.027778in}{0.000000in}}%
\pgfpathcurveto{\pgfqpoint{0.027778in}{0.007367in}}{\pgfqpoint{0.024851in}{0.014433in}}{\pgfqpoint{0.019642in}{0.019642in}}%
\pgfpathcurveto{\pgfqpoint{0.014433in}{0.024851in}}{\pgfqpoint{0.007367in}{0.027778in}}{\pgfqpoint{0.000000in}{0.027778in}}%
\pgfpathcurveto{\pgfqpoint{-0.007367in}{0.027778in}}{\pgfqpoint{-0.014433in}{0.024851in}}{\pgfqpoint{-0.019642in}{0.019642in}}%
\pgfpathcurveto{\pgfqpoint{-0.024851in}{0.014433in}}{\pgfqpoint{-0.027778in}{0.007367in}}{\pgfqpoint{-0.027778in}{0.000000in}}%
\pgfpathcurveto{\pgfqpoint{-0.027778in}{-0.007367in}}{\pgfqpoint{-0.024851in}{-0.014433in}}{\pgfqpoint{-0.019642in}{-0.019642in}}%
\pgfpathcurveto{\pgfqpoint{-0.014433in}{-0.024851in}}{\pgfqpoint{-0.007367in}{-0.027778in}}{\pgfqpoint{0.000000in}{-0.027778in}}%
\pgfpathlineto{\pgfqpoint{0.000000in}{-0.027778in}}%
\pgfpathclose%
\pgfusepath{stroke,fill}%
}%
\begin{pgfscope}%
\pgfsys@transformshift{0.544070in}{1.699175in}%
\pgfsys@useobject{currentmarker}{}%
\end{pgfscope}%
\begin{pgfscope}%
\pgfsys@transformshift{1.061390in}{1.704246in}%
\pgfsys@useobject{currentmarker}{}%
\end{pgfscope}%
\begin{pgfscope}%
\pgfsys@transformshift{1.578710in}{1.791322in}%
\pgfsys@useobject{currentmarker}{}%
\end{pgfscope}%
\begin{pgfscope}%
\pgfsys@transformshift{2.096030in}{1.961271in}%
\pgfsys@useobject{currentmarker}{}%
\end{pgfscope}%
\begin{pgfscope}%
\pgfsys@transformshift{2.613350in}{2.111045in}%
\pgfsys@useobject{currentmarker}{}%
\end{pgfscope}%
\begin{pgfscope}%
\pgfsys@transformshift{3.130670in}{2.176772in}%
\pgfsys@useobject{currentmarker}{}%
\end{pgfscope}%
\end{pgfscope}%
\begin{pgfscope}%
\pgfpathrectangle{\pgfqpoint{0.414740in}{0.362654in}}{\pgfqpoint{2.845260in}{2.016976in}}%
\pgfusepath{clip}%
\pgfsetrectcap%
\pgfsetroundjoin%
\pgfsetlinewidth{1.505625pt}%
\definecolor{currentstroke}{rgb}{0.773469,0.490168,0.965610}%
\pgfsetstrokecolor{currentstroke}%
\pgfsetdash{}{0pt}%
\pgfpathmoveto{\pgfqpoint{0.544070in}{1.570257in}}%
\pgfpathlineto{\pgfqpoint{1.061390in}{1.581530in}}%
\pgfpathlineto{\pgfqpoint{1.578710in}{1.690215in}}%
\pgfpathlineto{\pgfqpoint{2.096030in}{1.910456in}}%
\pgfpathlineto{\pgfqpoint{2.613350in}{2.101573in}}%
\pgfpathlineto{\pgfqpoint{3.130670in}{2.175053in}}%
\pgfusepath{stroke}%
\end{pgfscope}%
\begin{pgfscope}%
\pgfpathrectangle{\pgfqpoint{0.414740in}{0.362654in}}{\pgfqpoint{2.845260in}{2.016976in}}%
\pgfusepath{clip}%
\pgfsetbuttcap%
\pgfsetroundjoin%
\definecolor{currentfill}{rgb}{0.773469,0.490168,0.965610}%
\pgfsetfillcolor{currentfill}%
\pgfsetlinewidth{1.003750pt}%
\definecolor{currentstroke}{rgb}{0.773469,0.490168,0.965610}%
\pgfsetstrokecolor{currentstroke}%
\pgfsetdash{}{0pt}%
\pgfsys@defobject{currentmarker}{\pgfqpoint{-0.027778in}{-0.027778in}}{\pgfqpoint{0.027778in}{0.027778in}}{%
\pgfpathmoveto{\pgfqpoint{0.000000in}{-0.027778in}}%
\pgfpathcurveto{\pgfqpoint{0.007367in}{-0.027778in}}{\pgfqpoint{0.014433in}{-0.024851in}}{\pgfqpoint{0.019642in}{-0.019642in}}%
\pgfpathcurveto{\pgfqpoint{0.024851in}{-0.014433in}}{\pgfqpoint{0.027778in}{-0.007367in}}{\pgfqpoint{0.027778in}{0.000000in}}%
\pgfpathcurveto{\pgfqpoint{0.027778in}{0.007367in}}{\pgfqpoint{0.024851in}{0.014433in}}{\pgfqpoint{0.019642in}{0.019642in}}%
\pgfpathcurveto{\pgfqpoint{0.014433in}{0.024851in}}{\pgfqpoint{0.007367in}{0.027778in}}{\pgfqpoint{0.000000in}{0.027778in}}%
\pgfpathcurveto{\pgfqpoint{-0.007367in}{0.027778in}}{\pgfqpoint{-0.014433in}{0.024851in}}{\pgfqpoint{-0.019642in}{0.019642in}}%
\pgfpathcurveto{\pgfqpoint{-0.024851in}{0.014433in}}{\pgfqpoint{-0.027778in}{0.007367in}}{\pgfqpoint{-0.027778in}{0.000000in}}%
\pgfpathcurveto{\pgfqpoint{-0.027778in}{-0.007367in}}{\pgfqpoint{-0.024851in}{-0.014433in}}{\pgfqpoint{-0.019642in}{-0.019642in}}%
\pgfpathcurveto{\pgfqpoint{-0.014433in}{-0.024851in}}{\pgfqpoint{-0.007367in}{-0.027778in}}{\pgfqpoint{0.000000in}{-0.027778in}}%
\pgfpathlineto{\pgfqpoint{0.000000in}{-0.027778in}}%
\pgfpathclose%
\pgfusepath{stroke,fill}%
}%
\begin{pgfscope}%
\pgfsys@transformshift{0.544070in}{1.570257in}%
\pgfsys@useobject{currentmarker}{}%
\end{pgfscope}%
\begin{pgfscope}%
\pgfsys@transformshift{1.061390in}{1.581530in}%
\pgfsys@useobject{currentmarker}{}%
\end{pgfscope}%
\begin{pgfscope}%
\pgfsys@transformshift{1.578710in}{1.690215in}%
\pgfsys@useobject{currentmarker}{}%
\end{pgfscope}%
\begin{pgfscope}%
\pgfsys@transformshift{2.096030in}{1.910456in}%
\pgfsys@useobject{currentmarker}{}%
\end{pgfscope}%
\begin{pgfscope}%
\pgfsys@transformshift{2.613350in}{2.101573in}%
\pgfsys@useobject{currentmarker}{}%
\end{pgfscope}%
\begin{pgfscope}%
\pgfsys@transformshift{3.130670in}{2.175053in}%
\pgfsys@useobject{currentmarker}{}%
\end{pgfscope}%
\end{pgfscope}%
\begin{pgfscope}%
\pgfpathrectangle{\pgfqpoint{0.414740in}{0.362654in}}{\pgfqpoint{2.845260in}{2.016976in}}%
\pgfusepath{clip}%
\pgfsetrectcap%
\pgfsetroundjoin%
\pgfsetlinewidth{1.505625pt}%
\definecolor{currentstroke}{rgb}{0.196875,0.665670,0.761022}%
\pgfsetstrokecolor{currentstroke}%
\pgfsetdash{}{0pt}%
\pgfpathmoveto{\pgfqpoint{0.544070in}{1.441425in}}%
\pgfpathlineto{\pgfqpoint{1.061390in}{1.458848in}}%
\pgfpathlineto{\pgfqpoint{1.578710in}{1.583628in}}%
\pgfpathlineto{\pgfqpoint{2.096030in}{1.864637in}}%
\pgfpathlineto{\pgfqpoint{2.613350in}{2.094356in}}%
\pgfpathlineto{\pgfqpoint{3.130670in}{2.175425in}}%
\pgfusepath{stroke}%
\end{pgfscope}%
\begin{pgfscope}%
\pgfpathrectangle{\pgfqpoint{0.414740in}{0.362654in}}{\pgfqpoint{2.845260in}{2.016976in}}%
\pgfusepath{clip}%
\pgfsetbuttcap%
\pgfsetroundjoin%
\definecolor{currentfill}{rgb}{0.196875,0.665670,0.761022}%
\pgfsetfillcolor{currentfill}%
\pgfsetlinewidth{1.003750pt}%
\definecolor{currentstroke}{rgb}{0.196875,0.665670,0.761022}%
\pgfsetstrokecolor{currentstroke}%
\pgfsetdash{}{0pt}%
\pgfsys@defobject{currentmarker}{\pgfqpoint{-0.027778in}{-0.027778in}}{\pgfqpoint{0.027778in}{0.027778in}}{%
\pgfpathmoveto{\pgfqpoint{0.000000in}{-0.027778in}}%
\pgfpathcurveto{\pgfqpoint{0.007367in}{-0.027778in}}{\pgfqpoint{0.014433in}{-0.024851in}}{\pgfqpoint{0.019642in}{-0.019642in}}%
\pgfpathcurveto{\pgfqpoint{0.024851in}{-0.014433in}}{\pgfqpoint{0.027778in}{-0.007367in}}{\pgfqpoint{0.027778in}{0.000000in}}%
\pgfpathcurveto{\pgfqpoint{0.027778in}{0.007367in}}{\pgfqpoint{0.024851in}{0.014433in}}{\pgfqpoint{0.019642in}{0.019642in}}%
\pgfpathcurveto{\pgfqpoint{0.014433in}{0.024851in}}{\pgfqpoint{0.007367in}{0.027778in}}{\pgfqpoint{0.000000in}{0.027778in}}%
\pgfpathcurveto{\pgfqpoint{-0.007367in}{0.027778in}}{\pgfqpoint{-0.014433in}{0.024851in}}{\pgfqpoint{-0.019642in}{0.019642in}}%
\pgfpathcurveto{\pgfqpoint{-0.024851in}{0.014433in}}{\pgfqpoint{-0.027778in}{0.007367in}}{\pgfqpoint{-0.027778in}{0.000000in}}%
\pgfpathcurveto{\pgfqpoint{-0.027778in}{-0.007367in}}{\pgfqpoint{-0.024851in}{-0.014433in}}{\pgfqpoint{-0.019642in}{-0.019642in}}%
\pgfpathcurveto{\pgfqpoint{-0.014433in}{-0.024851in}}{\pgfqpoint{-0.007367in}{-0.027778in}}{\pgfqpoint{0.000000in}{-0.027778in}}%
\pgfpathlineto{\pgfqpoint{0.000000in}{-0.027778in}}%
\pgfpathclose%
\pgfusepath{stroke,fill}%
}%
\begin{pgfscope}%
\pgfsys@transformshift{0.544070in}{1.441425in}%
\pgfsys@useobject{currentmarker}{}%
\end{pgfscope}%
\begin{pgfscope}%
\pgfsys@transformshift{1.061390in}{1.458848in}%
\pgfsys@useobject{currentmarker}{}%
\end{pgfscope}%
\begin{pgfscope}%
\pgfsys@transformshift{1.578710in}{1.583628in}%
\pgfsys@useobject{currentmarker}{}%
\end{pgfscope}%
\begin{pgfscope}%
\pgfsys@transformshift{2.096030in}{1.864637in}%
\pgfsys@useobject{currentmarker}{}%
\end{pgfscope}%
\begin{pgfscope}%
\pgfsys@transformshift{2.613350in}{2.094356in}%
\pgfsys@useobject{currentmarker}{}%
\end{pgfscope}%
\begin{pgfscope}%
\pgfsys@transformshift{3.130670in}{2.175425in}%
\pgfsys@useobject{currentmarker}{}%
\end{pgfscope}%
\end{pgfscope}%
\begin{pgfscope}%
\pgfpathrectangle{\pgfqpoint{0.414740in}{0.362654in}}{\pgfqpoint{2.845260in}{2.016976in}}%
\pgfusepath{clip}%
\pgfsetrectcap%
\pgfsetroundjoin%
\pgfsetlinewidth{1.505625pt}%
\definecolor{currentstroke}{rgb}{0.179372,0.694079,0.447971}%
\pgfsetstrokecolor{currentstroke}%
\pgfsetdash{}{0pt}%
\pgfpathmoveto{\pgfqpoint{0.544070in}{1.328198in}}%
\pgfpathlineto{\pgfqpoint{1.061390in}{1.355706in}}%
\pgfpathlineto{\pgfqpoint{1.578710in}{1.498877in}}%
\pgfpathlineto{\pgfqpoint{2.096030in}{1.831401in}}%
\pgfpathlineto{\pgfqpoint{2.613350in}{2.089964in}}%
\pgfpathlineto{\pgfqpoint{3.130670in}{2.176674in}}%
\pgfusepath{stroke}%
\end{pgfscope}%
\begin{pgfscope}%
\pgfpathrectangle{\pgfqpoint{0.414740in}{0.362654in}}{\pgfqpoint{2.845260in}{2.016976in}}%
\pgfusepath{clip}%
\pgfsetbuttcap%
\pgfsetroundjoin%
\definecolor{currentfill}{rgb}{0.179372,0.694079,0.447971}%
\pgfsetfillcolor{currentfill}%
\pgfsetlinewidth{1.003750pt}%
\definecolor{currentstroke}{rgb}{0.179372,0.694079,0.447971}%
\pgfsetstrokecolor{currentstroke}%
\pgfsetdash{}{0pt}%
\pgfsys@defobject{currentmarker}{\pgfqpoint{-0.027778in}{-0.027778in}}{\pgfqpoint{0.027778in}{0.027778in}}{%
\pgfpathmoveto{\pgfqpoint{0.000000in}{-0.027778in}}%
\pgfpathcurveto{\pgfqpoint{0.007367in}{-0.027778in}}{\pgfqpoint{0.014433in}{-0.024851in}}{\pgfqpoint{0.019642in}{-0.019642in}}%
\pgfpathcurveto{\pgfqpoint{0.024851in}{-0.014433in}}{\pgfqpoint{0.027778in}{-0.007367in}}{\pgfqpoint{0.027778in}{0.000000in}}%
\pgfpathcurveto{\pgfqpoint{0.027778in}{0.007367in}}{\pgfqpoint{0.024851in}{0.014433in}}{\pgfqpoint{0.019642in}{0.019642in}}%
\pgfpathcurveto{\pgfqpoint{0.014433in}{0.024851in}}{\pgfqpoint{0.007367in}{0.027778in}}{\pgfqpoint{0.000000in}{0.027778in}}%
\pgfpathcurveto{\pgfqpoint{-0.007367in}{0.027778in}}{\pgfqpoint{-0.014433in}{0.024851in}}{\pgfqpoint{-0.019642in}{0.019642in}}%
\pgfpathcurveto{\pgfqpoint{-0.024851in}{0.014433in}}{\pgfqpoint{-0.027778in}{0.007367in}}{\pgfqpoint{-0.027778in}{0.000000in}}%
\pgfpathcurveto{\pgfqpoint{-0.027778in}{-0.007367in}}{\pgfqpoint{-0.024851in}{-0.014433in}}{\pgfqpoint{-0.019642in}{-0.019642in}}%
\pgfpathcurveto{\pgfqpoint{-0.014433in}{-0.024851in}}{\pgfqpoint{-0.007367in}{-0.027778in}}{\pgfqpoint{0.000000in}{-0.027778in}}%
\pgfpathlineto{\pgfqpoint{0.000000in}{-0.027778in}}%
\pgfpathclose%
\pgfusepath{stroke,fill}%
}%
\begin{pgfscope}%
\pgfsys@transformshift{0.544070in}{1.328198in}%
\pgfsys@useobject{currentmarker}{}%
\end{pgfscope}%
\begin{pgfscope}%
\pgfsys@transformshift{1.061390in}{1.355706in}%
\pgfsys@useobject{currentmarker}{}%
\end{pgfscope}%
\begin{pgfscope}%
\pgfsys@transformshift{1.578710in}{1.498877in}%
\pgfsys@useobject{currentmarker}{}%
\end{pgfscope}%
\begin{pgfscope}%
\pgfsys@transformshift{2.096030in}{1.831401in}%
\pgfsys@useobject{currentmarker}{}%
\end{pgfscope}%
\begin{pgfscope}%
\pgfsys@transformshift{2.613350in}{2.089964in}%
\pgfsys@useobject{currentmarker}{}%
\end{pgfscope}%
\begin{pgfscope}%
\pgfsys@transformshift{3.130670in}{2.176674in}%
\pgfsys@useobject{currentmarker}{}%
\end{pgfscope}%
\end{pgfscope}%
\begin{pgfscope}%
\pgfpathrectangle{\pgfqpoint{0.414740in}{0.362654in}}{\pgfqpoint{2.845260in}{2.016976in}}%
\pgfusepath{clip}%
\pgfsetbuttcap%
\pgfsetroundjoin%
\pgfsetlinewidth{1.505625pt}%
\definecolor{currentstroke}{rgb}{0.000000,0.000000,0.000000}%
\pgfsetstrokecolor{currentstroke}%
\pgfsetdash{{5.550000pt}{2.400000pt}}{0.000000pt}%
\pgfpathmoveto{\pgfqpoint{0.544070in}{0.880872in}}%
\pgfpathlineto{\pgfqpoint{1.061390in}{0.641245in}}%
\pgfpathlineto{\pgfqpoint{1.578710in}{0.454335in}}%
\pgfpathlineto{\pgfqpoint{2.096030in}{0.458471in}}%
\pgfpathlineto{\pgfqpoint{2.613350in}{0.487430in}}%
\pgfpathlineto{\pgfqpoint{3.130670in}{0.494069in}}%
\pgfusepath{stroke}%
\end{pgfscope}%
\begin{pgfscope}%
\pgfpathrectangle{\pgfqpoint{0.414740in}{0.362654in}}{\pgfqpoint{2.845260in}{2.016976in}}%
\pgfusepath{clip}%
\pgfsetbuttcap%
\pgfsetmiterjoin%
\definecolor{currentfill}{rgb}{0.000000,0.000000,0.000000}%
\pgfsetfillcolor{currentfill}%
\pgfsetlinewidth{1.003750pt}%
\definecolor{currentstroke}{rgb}{0.000000,0.000000,0.000000}%
\pgfsetstrokecolor{currentstroke}%
\pgfsetdash{}{0pt}%
\pgfsys@defobject{currentmarker}{\pgfqpoint{-0.027778in}{-0.027778in}}{\pgfqpoint{0.027778in}{0.027778in}}{%
\pgfpathmoveto{\pgfqpoint{-0.027778in}{-0.027778in}}%
\pgfpathlineto{\pgfqpoint{0.027778in}{-0.027778in}}%
\pgfpathlineto{\pgfqpoint{0.027778in}{0.027778in}}%
\pgfpathlineto{\pgfqpoint{-0.027778in}{0.027778in}}%
\pgfpathlineto{\pgfqpoint{-0.027778in}{-0.027778in}}%
\pgfpathclose%
\pgfusepath{stroke,fill}%
}%
\begin{pgfscope}%
\pgfsys@transformshift{0.544070in}{0.880872in}%
\pgfsys@useobject{currentmarker}{}%
\end{pgfscope}%
\begin{pgfscope}%
\pgfsys@transformshift{1.061390in}{0.641245in}%
\pgfsys@useobject{currentmarker}{}%
\end{pgfscope}%
\begin{pgfscope}%
\pgfsys@transformshift{1.578710in}{0.454335in}%
\pgfsys@useobject{currentmarker}{}%
\end{pgfscope}%
\begin{pgfscope}%
\pgfsys@transformshift{2.096030in}{0.458471in}%
\pgfsys@useobject{currentmarker}{}%
\end{pgfscope}%
\begin{pgfscope}%
\pgfsys@transformshift{2.613350in}{0.487430in}%
\pgfsys@useobject{currentmarker}{}%
\end{pgfscope}%
\begin{pgfscope}%
\pgfsys@transformshift{3.130670in}{0.494069in}%
\pgfsys@useobject{currentmarker}{}%
\end{pgfscope}%
\end{pgfscope}%
\begin{pgfscope}%
\pgfpathrectangle{\pgfqpoint{0.414740in}{0.362654in}}{\pgfqpoint{2.845260in}{2.016976in}}%
\pgfusepath{clip}%
\pgfsetbuttcap%
\pgfsetroundjoin%
\pgfsetlinewidth{1.505625pt}%
\definecolor{currentstroke}{rgb}{0.000000,0.000000,0.000000}%
\pgfsetstrokecolor{currentstroke}%
\pgfsetdash{{5.550000pt}{2.400000pt}}{0.000000pt}%
\pgfpathmoveto{\pgfqpoint{0.414740in}{2.287949in}}%
\pgfpathlineto{\pgfqpoint{3.260000in}{2.287949in}}%
\pgfusepath{stroke}%
\end{pgfscope}%
\begin{pgfscope}%
\pgfsetrectcap%
\pgfsetmiterjoin%
\pgfsetlinewidth{0.803000pt}%
\definecolor{currentstroke}{rgb}{0.000000,0.000000,0.000000}%
\pgfsetstrokecolor{currentstroke}%
\pgfsetdash{}{0pt}%
\pgfpathmoveto{\pgfqpoint{0.414740in}{0.362654in}}%
\pgfpathlineto{\pgfqpoint{0.414740in}{2.379630in}}%
\pgfusepath{stroke}%
\end{pgfscope}%
\begin{pgfscope}%
\pgfsetrectcap%
\pgfsetmiterjoin%
\pgfsetlinewidth{0.803000pt}%
\definecolor{currentstroke}{rgb}{0.000000,0.000000,0.000000}%
\pgfsetstrokecolor{currentstroke}%
\pgfsetdash{}{0pt}%
\pgfpathmoveto{\pgfqpoint{3.260000in}{0.362654in}}%
\pgfpathlineto{\pgfqpoint{3.260000in}{2.379630in}}%
\pgfusepath{stroke}%
\end{pgfscope}%
\begin{pgfscope}%
\pgfsetrectcap%
\pgfsetmiterjoin%
\pgfsetlinewidth{0.803000pt}%
\definecolor{currentstroke}{rgb}{0.000000,0.000000,0.000000}%
\pgfsetstrokecolor{currentstroke}%
\pgfsetdash{}{0pt}%
\pgfpathmoveto{\pgfqpoint{0.414740in}{0.362654in}}%
\pgfpathlineto{\pgfqpoint{3.260000in}{0.362654in}}%
\pgfusepath{stroke}%
\end{pgfscope}%
\begin{pgfscope}%
\pgfsetrectcap%
\pgfsetmiterjoin%
\pgfsetlinewidth{0.803000pt}%
\definecolor{currentstroke}{rgb}{0.000000,0.000000,0.000000}%
\pgfsetstrokecolor{currentstroke}%
\pgfsetdash{}{0pt}%
\pgfpathmoveto{\pgfqpoint{0.414740in}{2.379630in}}%
\pgfpathlineto{\pgfqpoint{3.260000in}{2.379630in}}%
\pgfusepath{stroke}%
\end{pgfscope}%
\begin{pgfscope}%
\pgfsetbuttcap%
\pgfsetmiterjoin%
\definecolor{currentfill}{rgb}{1.000000,1.000000,1.000000}%
\pgfsetfillcolor{currentfill}%
\pgfsetfillopacity{0.800000}%
\pgfsetlinewidth{1.003750pt}%
\definecolor{currentstroke}{rgb}{0.800000,0.800000,0.800000}%
\pgfsetstrokecolor{currentstroke}%
\pgfsetstrokeopacity{0.800000}%
\pgfsetdash{}{0pt}%
\pgfpathmoveto{\pgfqpoint{2.187537in}{0.687963in}}%
\pgfpathlineto{\pgfqpoint{3.182222in}{0.687963in}}%
\pgfpathquadraticcurveto{\pgfqpoint{3.204444in}{0.687963in}}{\pgfqpoint{3.204444in}{0.710185in}}%
\pgfpathlineto{\pgfqpoint{3.204444in}{1.628704in}}%
\pgfpathquadraticcurveto{\pgfqpoint{3.204444in}{1.650926in}}{\pgfqpoint{3.182222in}{1.650926in}}%
\pgfpathlineto{\pgfqpoint{2.187537in}{1.650926in}}%
\pgfpathquadraticcurveto{\pgfqpoint{2.165315in}{1.650926in}}{\pgfqpoint{2.165315in}{1.628704in}}%
\pgfpathlineto{\pgfqpoint{2.165315in}{0.710185in}}%
\pgfpathquadraticcurveto{\pgfqpoint{2.165315in}{0.687963in}}{\pgfqpoint{2.187537in}{0.687963in}}%
\pgfpathlineto{\pgfqpoint{2.187537in}{0.687963in}}%
\pgfpathclose%
\pgfusepath{stroke,fill}%
\end{pgfscope}%
\begin{pgfscope}%
\pgfsetrectcap%
\pgfsetroundjoin%
\pgfsetlinewidth{1.505625pt}%
\definecolor{currentstroke}{rgb}{0.973355,0.432701,0.565100}%
\pgfsetstrokecolor{currentstroke}%
\pgfsetdash{}{0pt}%
\pgfpathmoveto{\pgfqpoint{2.209759in}{1.567593in}}%
\pgfpathlineto{\pgfqpoint{2.320870in}{1.567593in}}%
\pgfpathlineto{\pgfqpoint{2.431982in}{1.567593in}}%
\pgfusepath{stroke}%
\end{pgfscope}%
\begin{pgfscope}%
\pgfsetbuttcap%
\pgfsetroundjoin%
\definecolor{currentfill}{rgb}{0.973355,0.432701,0.565100}%
\pgfsetfillcolor{currentfill}%
\pgfsetlinewidth{1.003750pt}%
\definecolor{currentstroke}{rgb}{0.973355,0.432701,0.565100}%
\pgfsetstrokecolor{currentstroke}%
\pgfsetdash{}{0pt}%
\pgfsys@defobject{currentmarker}{\pgfqpoint{-0.027778in}{-0.027778in}}{\pgfqpoint{0.027778in}{0.027778in}}{%
\pgfpathmoveto{\pgfqpoint{0.000000in}{-0.027778in}}%
\pgfpathcurveto{\pgfqpoint{0.007367in}{-0.027778in}}{\pgfqpoint{0.014433in}{-0.024851in}}{\pgfqpoint{0.019642in}{-0.019642in}}%
\pgfpathcurveto{\pgfqpoint{0.024851in}{-0.014433in}}{\pgfqpoint{0.027778in}{-0.007367in}}{\pgfqpoint{0.027778in}{0.000000in}}%
\pgfpathcurveto{\pgfqpoint{0.027778in}{0.007367in}}{\pgfqpoint{0.024851in}{0.014433in}}{\pgfqpoint{0.019642in}{0.019642in}}%
\pgfpathcurveto{\pgfqpoint{0.014433in}{0.024851in}}{\pgfqpoint{0.007367in}{0.027778in}}{\pgfqpoint{0.000000in}{0.027778in}}%
\pgfpathcurveto{\pgfqpoint{-0.007367in}{0.027778in}}{\pgfqpoint{-0.014433in}{0.024851in}}{\pgfqpoint{-0.019642in}{0.019642in}}%
\pgfpathcurveto{\pgfqpoint{-0.024851in}{0.014433in}}{\pgfqpoint{-0.027778in}{0.007367in}}{\pgfqpoint{-0.027778in}{0.000000in}}%
\pgfpathcurveto{\pgfqpoint{-0.027778in}{-0.007367in}}{\pgfqpoint{-0.024851in}{-0.014433in}}{\pgfqpoint{-0.019642in}{-0.019642in}}%
\pgfpathcurveto{\pgfqpoint{-0.014433in}{-0.024851in}}{\pgfqpoint{-0.007367in}{-0.027778in}}{\pgfqpoint{0.000000in}{-0.027778in}}%
\pgfpathlineto{\pgfqpoint{0.000000in}{-0.027778in}}%
\pgfpathclose%
\pgfusepath{stroke,fill}%
}%
\begin{pgfscope}%
\pgfsys@transformshift{2.320870in}{1.567593in}%
\pgfsys@useobject{currentmarker}{}%
\end{pgfscope}%
\end{pgfscope}%
\begin{pgfscope}%
\definecolor{textcolor}{rgb}{0.000000,0.000000,0.000000}%
\pgfsetstrokecolor{textcolor}%
\pgfsetfillcolor{textcolor}%
\pgftext[x=2.520870in,y=1.528704in,left,base]{\color{textcolor}{\rmfamily\fontsize{8.000000}{9.600000}\selectfont\catcode`\^=\active\def^{\ifmmode\sp\else\^{}\fi}\catcode`\%=\active\def%{\%}$M = 64$}}%
\end{pgfscope}%
\begin{pgfscope}%
\pgfsetrectcap%
\pgfsetroundjoin%
\pgfsetlinewidth{1.505625pt}%
\definecolor{currentstroke}{rgb}{0.773469,0.490168,0.965610}%
\pgfsetstrokecolor{currentstroke}%
\pgfsetdash{}{0pt}%
\pgfpathmoveto{\pgfqpoint{2.209759in}{1.412655in}}%
\pgfpathlineto{\pgfqpoint{2.320870in}{1.412655in}}%
\pgfpathlineto{\pgfqpoint{2.431982in}{1.412655in}}%
\pgfusepath{stroke}%
\end{pgfscope}%
\begin{pgfscope}%
\pgfsetbuttcap%
\pgfsetroundjoin%
\definecolor{currentfill}{rgb}{0.773469,0.490168,0.965610}%
\pgfsetfillcolor{currentfill}%
\pgfsetlinewidth{1.003750pt}%
\definecolor{currentstroke}{rgb}{0.773469,0.490168,0.965610}%
\pgfsetstrokecolor{currentstroke}%
\pgfsetdash{}{0pt}%
\pgfsys@defobject{currentmarker}{\pgfqpoint{-0.027778in}{-0.027778in}}{\pgfqpoint{0.027778in}{0.027778in}}{%
\pgfpathmoveto{\pgfqpoint{0.000000in}{-0.027778in}}%
\pgfpathcurveto{\pgfqpoint{0.007367in}{-0.027778in}}{\pgfqpoint{0.014433in}{-0.024851in}}{\pgfqpoint{0.019642in}{-0.019642in}}%
\pgfpathcurveto{\pgfqpoint{0.024851in}{-0.014433in}}{\pgfqpoint{0.027778in}{-0.007367in}}{\pgfqpoint{0.027778in}{0.000000in}}%
\pgfpathcurveto{\pgfqpoint{0.027778in}{0.007367in}}{\pgfqpoint{0.024851in}{0.014433in}}{\pgfqpoint{0.019642in}{0.019642in}}%
\pgfpathcurveto{\pgfqpoint{0.014433in}{0.024851in}}{\pgfqpoint{0.007367in}{0.027778in}}{\pgfqpoint{0.000000in}{0.027778in}}%
\pgfpathcurveto{\pgfqpoint{-0.007367in}{0.027778in}}{\pgfqpoint{-0.014433in}{0.024851in}}{\pgfqpoint{-0.019642in}{0.019642in}}%
\pgfpathcurveto{\pgfqpoint{-0.024851in}{0.014433in}}{\pgfqpoint{-0.027778in}{0.007367in}}{\pgfqpoint{-0.027778in}{0.000000in}}%
\pgfpathcurveto{\pgfqpoint{-0.027778in}{-0.007367in}}{\pgfqpoint{-0.024851in}{-0.014433in}}{\pgfqpoint{-0.019642in}{-0.019642in}}%
\pgfpathcurveto{\pgfqpoint{-0.014433in}{-0.024851in}}{\pgfqpoint{-0.007367in}{-0.027778in}}{\pgfqpoint{0.000000in}{-0.027778in}}%
\pgfpathlineto{\pgfqpoint{0.000000in}{-0.027778in}}%
\pgfpathclose%
\pgfusepath{stroke,fill}%
}%
\begin{pgfscope}%
\pgfsys@transformshift{2.320870in}{1.412655in}%
\pgfsys@useobject{currentmarker}{}%
\end{pgfscope}%
\end{pgfscope}%
\begin{pgfscope}%
\definecolor{textcolor}{rgb}{0.000000,0.000000,0.000000}%
\pgfsetstrokecolor{textcolor}%
\pgfsetfillcolor{textcolor}%
\pgftext[x=2.520870in,y=1.373766in,left,base]{\color{textcolor}{\rmfamily\fontsize{8.000000}{9.600000}\selectfont\catcode`\^=\active\def^{\ifmmode\sp\else\^{}\fi}\catcode`\%=\active\def%{\%}$M = 32$}}%
\end{pgfscope}%
\begin{pgfscope}%
\pgfsetrectcap%
\pgfsetroundjoin%
\pgfsetlinewidth{1.505625pt}%
\definecolor{currentstroke}{rgb}{0.196875,0.665670,0.761022}%
\pgfsetstrokecolor{currentstroke}%
\pgfsetdash{}{0pt}%
\pgfpathmoveto{\pgfqpoint{2.209759in}{1.257716in}}%
\pgfpathlineto{\pgfqpoint{2.320870in}{1.257716in}}%
\pgfpathlineto{\pgfqpoint{2.431982in}{1.257716in}}%
\pgfusepath{stroke}%
\end{pgfscope}%
\begin{pgfscope}%
\pgfsetbuttcap%
\pgfsetroundjoin%
\definecolor{currentfill}{rgb}{0.196875,0.665670,0.761022}%
\pgfsetfillcolor{currentfill}%
\pgfsetlinewidth{1.003750pt}%
\definecolor{currentstroke}{rgb}{0.196875,0.665670,0.761022}%
\pgfsetstrokecolor{currentstroke}%
\pgfsetdash{}{0pt}%
\pgfsys@defobject{currentmarker}{\pgfqpoint{-0.027778in}{-0.027778in}}{\pgfqpoint{0.027778in}{0.027778in}}{%
\pgfpathmoveto{\pgfqpoint{0.000000in}{-0.027778in}}%
\pgfpathcurveto{\pgfqpoint{0.007367in}{-0.027778in}}{\pgfqpoint{0.014433in}{-0.024851in}}{\pgfqpoint{0.019642in}{-0.019642in}}%
\pgfpathcurveto{\pgfqpoint{0.024851in}{-0.014433in}}{\pgfqpoint{0.027778in}{-0.007367in}}{\pgfqpoint{0.027778in}{0.000000in}}%
\pgfpathcurveto{\pgfqpoint{0.027778in}{0.007367in}}{\pgfqpoint{0.024851in}{0.014433in}}{\pgfqpoint{0.019642in}{0.019642in}}%
\pgfpathcurveto{\pgfqpoint{0.014433in}{0.024851in}}{\pgfqpoint{0.007367in}{0.027778in}}{\pgfqpoint{0.000000in}{0.027778in}}%
\pgfpathcurveto{\pgfqpoint{-0.007367in}{0.027778in}}{\pgfqpoint{-0.014433in}{0.024851in}}{\pgfqpoint{-0.019642in}{0.019642in}}%
\pgfpathcurveto{\pgfqpoint{-0.024851in}{0.014433in}}{\pgfqpoint{-0.027778in}{0.007367in}}{\pgfqpoint{-0.027778in}{0.000000in}}%
\pgfpathcurveto{\pgfqpoint{-0.027778in}{-0.007367in}}{\pgfqpoint{-0.024851in}{-0.014433in}}{\pgfqpoint{-0.019642in}{-0.019642in}}%
\pgfpathcurveto{\pgfqpoint{-0.014433in}{-0.024851in}}{\pgfqpoint{-0.007367in}{-0.027778in}}{\pgfqpoint{0.000000in}{-0.027778in}}%
\pgfpathlineto{\pgfqpoint{0.000000in}{-0.027778in}}%
\pgfpathclose%
\pgfusepath{stroke,fill}%
}%
\begin{pgfscope}%
\pgfsys@transformshift{2.320870in}{1.257716in}%
\pgfsys@useobject{currentmarker}{}%
\end{pgfscope}%
\end{pgfscope}%
\begin{pgfscope}%
\definecolor{textcolor}{rgb}{0.000000,0.000000,0.000000}%
\pgfsetstrokecolor{textcolor}%
\pgfsetfillcolor{textcolor}%
\pgftext[x=2.520870in,y=1.218827in,left,base]{\color{textcolor}{\rmfamily\fontsize{8.000000}{9.600000}\selectfont\catcode`\^=\active\def^{\ifmmode\sp\else\^{}\fi}\catcode`\%=\active\def%{\%}$M = 16$}}%
\end{pgfscope}%
\begin{pgfscope}%
\pgfsetrectcap%
\pgfsetroundjoin%
\pgfsetlinewidth{1.505625pt}%
\definecolor{currentstroke}{rgb}{0.179372,0.694079,0.447971}%
\pgfsetstrokecolor{currentstroke}%
\pgfsetdash{}{0pt}%
\pgfpathmoveto{\pgfqpoint{2.209759in}{1.102778in}}%
\pgfpathlineto{\pgfqpoint{2.320870in}{1.102778in}}%
\pgfpathlineto{\pgfqpoint{2.431982in}{1.102778in}}%
\pgfusepath{stroke}%
\end{pgfscope}%
\begin{pgfscope}%
\pgfsetbuttcap%
\pgfsetroundjoin%
\definecolor{currentfill}{rgb}{0.179372,0.694079,0.447971}%
\pgfsetfillcolor{currentfill}%
\pgfsetlinewidth{1.003750pt}%
\definecolor{currentstroke}{rgb}{0.179372,0.694079,0.447971}%
\pgfsetstrokecolor{currentstroke}%
\pgfsetdash{}{0pt}%
\pgfsys@defobject{currentmarker}{\pgfqpoint{-0.027778in}{-0.027778in}}{\pgfqpoint{0.027778in}{0.027778in}}{%
\pgfpathmoveto{\pgfqpoint{0.000000in}{-0.027778in}}%
\pgfpathcurveto{\pgfqpoint{0.007367in}{-0.027778in}}{\pgfqpoint{0.014433in}{-0.024851in}}{\pgfqpoint{0.019642in}{-0.019642in}}%
\pgfpathcurveto{\pgfqpoint{0.024851in}{-0.014433in}}{\pgfqpoint{0.027778in}{-0.007367in}}{\pgfqpoint{0.027778in}{0.000000in}}%
\pgfpathcurveto{\pgfqpoint{0.027778in}{0.007367in}}{\pgfqpoint{0.024851in}{0.014433in}}{\pgfqpoint{0.019642in}{0.019642in}}%
\pgfpathcurveto{\pgfqpoint{0.014433in}{0.024851in}}{\pgfqpoint{0.007367in}{0.027778in}}{\pgfqpoint{0.000000in}{0.027778in}}%
\pgfpathcurveto{\pgfqpoint{-0.007367in}{0.027778in}}{\pgfqpoint{-0.014433in}{0.024851in}}{\pgfqpoint{-0.019642in}{0.019642in}}%
\pgfpathcurveto{\pgfqpoint{-0.024851in}{0.014433in}}{\pgfqpoint{-0.027778in}{0.007367in}}{\pgfqpoint{-0.027778in}{0.000000in}}%
\pgfpathcurveto{\pgfqpoint{-0.027778in}{-0.007367in}}{\pgfqpoint{-0.024851in}{-0.014433in}}{\pgfqpoint{-0.019642in}{-0.019642in}}%
\pgfpathcurveto{\pgfqpoint{-0.014433in}{-0.024851in}}{\pgfqpoint{-0.007367in}{-0.027778in}}{\pgfqpoint{0.000000in}{-0.027778in}}%
\pgfpathlineto{\pgfqpoint{0.000000in}{-0.027778in}}%
\pgfpathclose%
\pgfusepath{stroke,fill}%
}%
\begin{pgfscope}%
\pgfsys@transformshift{2.320870in}{1.102778in}%
\pgfsys@useobject{currentmarker}{}%
\end{pgfscope}%
\end{pgfscope}%
\begin{pgfscope}%
\definecolor{textcolor}{rgb}{0.000000,0.000000,0.000000}%
\pgfsetstrokecolor{textcolor}%
\pgfsetfillcolor{textcolor}%
\pgftext[x=2.520870in,y=1.063889in,left,base]{\color{textcolor}{\rmfamily\fontsize{8.000000}{9.600000}\selectfont\catcode`\^=\active\def^{\ifmmode\sp\else\^{}\fi}\catcode`\%=\active\def%{\%}$M = 8$}}%
\end{pgfscope}%
\begin{pgfscope}%
\pgfsetbuttcap%
\pgfsetroundjoin%
\pgfsetlinewidth{1.505625pt}%
\definecolor{currentstroke}{rgb}{0.000000,0.000000,0.000000}%
\pgfsetstrokecolor{currentstroke}%
\pgfsetdash{{5.550000pt}{2.400000pt}}{0.000000pt}%
\pgfpathmoveto{\pgfqpoint{2.209759in}{0.947840in}}%
\pgfpathlineto{\pgfqpoint{2.320870in}{0.947840in}}%
\pgfpathlineto{\pgfqpoint{2.431982in}{0.947840in}}%
\pgfusepath{stroke}%
\end{pgfscope}%
\begin{pgfscope}%
\pgfsetbuttcap%
\pgfsetmiterjoin%
\definecolor{currentfill}{rgb}{0.000000,0.000000,0.000000}%
\pgfsetfillcolor{currentfill}%
\pgfsetlinewidth{1.003750pt}%
\definecolor{currentstroke}{rgb}{0.000000,0.000000,0.000000}%
\pgfsetstrokecolor{currentstroke}%
\pgfsetdash{}{0pt}%
\pgfsys@defobject{currentmarker}{\pgfqpoint{-0.027778in}{-0.027778in}}{\pgfqpoint{0.027778in}{0.027778in}}{%
\pgfpathmoveto{\pgfqpoint{-0.027778in}{-0.027778in}}%
\pgfpathlineto{\pgfqpoint{0.027778in}{-0.027778in}}%
\pgfpathlineto{\pgfqpoint{0.027778in}{0.027778in}}%
\pgfpathlineto{\pgfqpoint{-0.027778in}{0.027778in}}%
\pgfpathlineto{\pgfqpoint{-0.027778in}{-0.027778in}}%
\pgfpathclose%
\pgfusepath{stroke,fill}%
}%
\begin{pgfscope}%
\pgfsys@transformshift{2.320870in}{0.947840in}%
\pgfsys@useobject{currentmarker}{}%
\end{pgfscope}%
\end{pgfscope}%
\begin{pgfscope}%
\definecolor{textcolor}{rgb}{0.000000,0.000000,0.000000}%
\pgfsetstrokecolor{textcolor}%
\pgfsetfillcolor{textcolor}%
\pgftext[x=2.520870in,y=0.908951in,left,base]{\color{textcolor}{\rmfamily\fontsize{8.000000}{9.600000}\selectfont\catcode`\^=\active\def^{\ifmmode\sp\else\^{}\fi}\catcode`\%=\active\def%{\%}MRT+MRC}}%
\end{pgfscope}%
\begin{pgfscope}%
\pgfsetbuttcap%
\pgfsetroundjoin%
\pgfsetlinewidth{1.505625pt}%
\definecolor{currentstroke}{rgb}{0.000000,0.000000,0.000000}%
\pgfsetstrokecolor{currentstroke}%
\pgfsetdash{{5.550000pt}{2.400000pt}}{0.000000pt}%
\pgfpathmoveto{\pgfqpoint{2.209759in}{0.792901in}}%
\pgfpathlineto{\pgfqpoint{2.320870in}{0.792901in}}%
\pgfpathlineto{\pgfqpoint{2.431982in}{0.792901in}}%
\pgfusepath{stroke}%
\end{pgfscope}%
\begin{pgfscope}%
\definecolor{textcolor}{rgb}{0.000000,0.000000,0.000000}%
\pgfsetstrokecolor{textcolor}%
\pgfsetfillcolor{textcolor}%
\pgftext[x=2.520870in,y=0.754012in,left,base]{\color{textcolor}{\rmfamily\fontsize{8.000000}{9.600000}\selectfont\catcode`\^=\active\def^{\ifmmode\sp\else\^{}\fi}\catcode`\%=\active\def%{\%}Capacity}}%
\end{pgfscope}%
\end{pgfpicture}%
\makeatother%
\endgroup%

%% file: sec-conclusion.tex
\section{Conclusion} \label{sec:conclusion}

This work develops a novel beam learning framework for full-duplex \BSs that aims to achieve high \sse while minimizing the number of probing beams required for \SI channel estimation.
This is realized by designing site-specific probing codebooks to implicitly estimate the \SI channel and then synthesizing the final \DL and \UL beams using a deep learning model.
Our approach is in stark contrast to many existing schemes, whose reliance on explicit estimation of the \SI channel would incur prohibitive overhead in most real-world settings.
We demonstrate that our approach can suppress \SI to near or below the noise floor and attain high \sse with only 8--64 probing measurements, even when the self-interference channel itself contains 256 unknown elements.
Numerical results using ray-tracing also illustrates our model's robustness to varying channel conditions.
Valuable future work may explore how to extend ideas herein to multi-user systems or may experimentally demonstrate beam learning schemes using actual wireless platforms.

%% file: sec-acknowledgement.tex
\section*{Acknowledgments}
This work used computational and storage services from the Hoffman2 Cluster, operated by the UCLA Office of Advanced Research Computing's Research Technology Group.

%% file: sec-bibliography.tex
\bibliographystyle{bibtex/IEEEtran}
\bibliography{bibtex/IEEEabrv,refs}